\documentclass[10pt,conference]{IEEEtran}
\IEEEoverridecommandlockouts

\usepackage{amsmath,amssymb,amsfonts}

\usepackage{booktabs}   
\usepackage{subcaption} 
\newsavebox{\mybox}

\usepackage{amsthm}
\usepackage{mdframed}
\usepackage{centernot}
\usepackage{comment}

\newcommand{\Aa}{\mathcal{A}}

\newcommand{\Ff}{\mathcal{F}}

\newcommand{\Tt}{\mathcal{T}}
\newcommand{\Hh}{\mathcal{H}}

\newcommand{\Pp}{\mathcal{P}}
\newcommand{\sPp}{\sem{\Pp}}

\newcommand{\Ss}{\Sigma}

\newcommand\newsubcap[1]{\phantomcaption%
       \caption*{\figurename~\thefigure.~(\thesubfigure) #1}}

\newcommand{\sem}[1]{ [ \! [ {#1}  ]  \! ]} 
\def\rmdef{\stackrel{\mbox{\rm {\tiny def}}}{=}} 

\newcommand\vx{\mathbf{x}}

\newcommand\vd[2]{d_{i, p}}
\newcommand\vy{\mathbf{y}}
\newcommand\vz{\mathbf{z}}

\newcommand{\set}[1]{\left\{ #1 \right\}}

\newcommand{\seq}[1]{\langle #1 \rangle}
\newcommand{\Nat}{\mathbb N}

\newcommand{\R}{\mathbb R}

\newcommand{\Real}{\R}
\newcommand{\Rplus}{\R_{\geq 0}}
\newcommand{\toolname}{\textsc{Fuchsia}\xspace}
\usepackage{tikz}

\newtheorem{definition}{Definition}[section]

\definecolor{gold}{rgb}{0.99,0.78,0.07}
\usetikzlibrary{arrows,shapes,snakes,automata,backgrounds,positioning,decorations.pathmorphing,
	decorations.markings,calc}

\tikzstyle{dtreenode}=[draw=blue!10!gray,rounded rectangle, minimum size=5mm,fill=blue!10!white]
\tikzstyle{dtreeleaf}=[draw=black!60,minimum width=1cm,minimum height=0.4cm,rectangle,fill=blue!50!white]
\tikzset{every loop/.style={looseness=7}}
\tikzset{
	gluon/.style={decorate,draw=black,
		decoration={coil,amplitude=1pt, segment length=5pt}}
}
\tikzset{
	gluon1/.style={decorate,draw=black,
		decoration={coil,amplitude=3pt, segment length=3pt}}
}
\tikzset{
	gluonew/.style={decorate,draw=black,
		decoration={coil,amplitude=1pt, segment length=2pt}}
}

\tikzset{bicolor/.style args={#1 and #2 and #3}{
		path picture={
			\tikzset{rounded corners=0}
			\fill [#1] (path picture bounding box.south west)
			rectangle
			($(path picture  bounding box.north west)!#3!(path picture bounding
			box.north east)$);
			\fill [#2]
			($(path picture bounding box.south west)!#3!(path picture bounding
			box.south east)$)
			rectangle (path picture bounding box.north east);
}}}

\tikzset{tricolor/.style args={#1 and #2 and #3 and #4 and #5}{
		path picture={
			\tikzset{rounded corners=0}
			\fill [#1] (path picture bounding box.south west)
			rectangle
			($(path picture  bounding box.north west)!#4!(path picture bounding
			box.north east)$);
			\fill [#2]
			($(path picture bounding box.south west)!#4!(path picture bounding
			box.south east)$)
			rectangle
			($(path picture  bounding box.north west)!#5!(path picture bounding
			box.north east)$);
			\fill [#3]
			($(path picture bounding box.south west)!#5!(path picture bounding
			box.south east)$)
			rectangle (path picture bounding box.north east);
}}}

\usepackage{tcolorbox}
\tcbuselibrary{listings,skins}

\lstdefinestyle{mystyle}{
  xleftmargin=0pt,
   basicstyle={\footnotesize\ttfamily},
   aboveskip=3mm,
   belowskip=3mm,
   keywordstyle=\bfseries,
   showstringspaces=false,
  escapechar=?,
  language=Java
}
\definecolor{code_indent}{HTML}{CCCCCC}

\newcommand{\indentrule}{\color{code_indent}\vrule\hspace{2pt}}

\newtcblisting{mylisting}[2][]{
  arc=0pt, outer arc=0pt,
  listing only,
  listing style=mystyle,
  title={\large #2},
  #1
}

 \definecolor{dkgreen}{rgb}{0,0.6,0}
 \definecolor{gray}{rgb}{0.5,0.5,0.5}
 \definecolor{mauve}{rgb}{0.58,0,0.82}
 \definecolor{violet}{rgb}{0.56, 0.0, 1.0}
\definecolor{aquamarine}{rgb}{0.5, 1.0, 0.83}
\definecolor{lightblue}{rgb}{0.5, 0.83, 1.0}
\definecolor{brightgreen}{rgb}{0.5, 1.0, 0.0}
\definecolor{brightgreen_2}{rgb}{0.2, 1.0, 0.0}
\definecolor{amber}{rgb}{1.0, 0.75, 0.0}

\definecolor{cadmiumgreen}{rgb}{0.0, 0.42, 0.24}
\definecolor{verde}{rgb}{0.25,0.5,0.35}
\definecolor{jpurple}{rgb}{0.5,0,0.35}
\definecolor{darkgreen}{rgb}{0.0, 0.2, 0.13}

\usepackage[ruled,vlined,linesnumbered,lined]{algorithm2e}
\usepackage{algpseudocode}

\usepackage{mathtools,xparse}

\newcommand*\circled[1]{\tikz[baseline=(char.base)]{
            \node[shape=circle,draw,inner sep=2pt] (char) {#1};}}
\PassOptionsToPackage{hyphens}{url}\usepackage{hyperref}

\begin{document}

\title{\scalebox{0.9}{Data-Driven Debugging for Functional Side Channels}}
\author{\IEEEauthorblockN{Saeid Tizpaz-Niari}
\IEEEauthorblockA{University of Colorado Boulder\\
saeid.tizpazniari@colorado.edu}
\and
\IEEEauthorblockN{Pavol {\v C}ern\'y}
\IEEEauthorblockA{TU Wien\\
pavol.cerny@tuwien.ac.at}
\and
\IEEEauthorblockN{Ashutosh Trivedi}
\IEEEauthorblockA{University of Colorado Boulder\\
ashutosh.trivedi@colorado.edu}}

\IEEEoverridecommandlockouts
\makeatletter\def\@IEEEpubidpullup{6.5\baselineskip}\makeatother
\IEEEpubid{\parbox{\columnwidth}{
    Network and Distributed Systems Security (NDSS) Symposium 2020\\
    23-26 February 2020, San Diego, CA, USA\\
    ISBN 1-891562-61-4\\
    https://dx.doi.org/10.14722/ndss.2020.24269\\
    www.ndss-symposium.org
}
\hspace{\columnsep}\makebox[\columnwidth]{}}

\maketitle

\begin{abstract}
Information leaks through side channels are a pervasive problem, even
in security-critical applications.
{\em Functional side channels} arise when an attacker knows that a secret
value of a server stays fixed for a certain time. Then, the attacker
can observe the server
executions on a sequence of different public inputs, each paired with the same
secret input.
Thus for each secret, the attacker observes a (partial) function from public
inputs to execution time, for instance, and she can compare these functions for
different secrets.

First, we introduce a notion of noninterference for functional side channels.
We focus on the case of noisy observations, where we demonstrate with
examples that there is a practical functional side channel in programs that
would be deemed information-leak-free or be underestimated using
the standard definition.
Second, we develop a framework and techniques for debugging programs for
functional side channels.
We extend evolutionary fuzzing techniques to generate
inputs that exploit functional dependencies of response times on public inputs.
We adapt existing results and algorithms in functional data analysis (such as
functional clustering) to model the functions and discover the existence of
side channels.
We use a functional extension of standard
decision tree learning to pinpoint the code fragments causing
a side channel if there is one.

We empirically evaluate the performance of our tool \toolname on a
series of micro-benchmarks, as well as on realistic Java programs.
On the set of micro-benchmark, we show that \toolname outperforms
the state-of-the-art techniques in detecting side-channel classes.
On the realistic programs, we show the scalability of \toolname in
analyzing functional side channels in Java programs
with thousands of methods. In addition, we show the usefulness of \toolname
in finding (and locating in code) side channels including a zero-day
vulnerability in Open Java Development Kit and another Java web server
vulnerability that was since fixed by the original developers.
\end{abstract}

\section{Introduction}
\label{sec:introduction}
Developers are careful to assure that eavesdroppers cannot easily access the
secrets by employing security practices such as encryption.
However, a side
channel might arise even if the transferred data is encrypted. The
side-channel eavesdroppers can infer the value of secret inputs
(or some of their properties) based on public inputs,
runtime observations, and the source code of the
program. An example is OnlineHealth service~\cite{CWWZ10}, where
the service leaks the conditions of patients through
side channels observable in the characteristics of network packets.

We consider {\em known-message} threats~\cite{kopf2009provably}
where the attacker knows the
value of public inputs as well as execution times when trying
to find out the secret. In this threat model, we consider the setting where the secret input stays fixed across a number of
interactions. This gives rise to {\em functional observations}: for a secret
input, we observe the program executions on a number of public inputs. For a secret input
$s$, we obtain a partial function $f_s$ from public inputs to runtime
observations. We focus on timing side channels, where the
attacker's observations are the execution time.

\noindent{\bf Functional side channels.} We adapt the classical definition of
noninterference to {\em functional side channels}, where two secret inputs $s$
and $t$ are indistinguishable for the attacker if the functions $f_s$ and $f_t$
are equal. However, in the presence of noise (a common situation for timing
measurements), we cannot require exact equality of functions. Instead, we
define two functional observations to be indistinguishable when they are
similar according to a notion of distance.
We demonstrate on a set of examples that it is critical to choose the
distance that represents functional observations,
otherwise, side channels might remain undetected or be underestimated.

\noindent{\bf Problem.} {\it Data-driven debugging focuses on
automatically discovering functional timing side channels,
and on pinpointing code regions that contribute to creating the side channels.}

\noindent{\bf Algorithms.}  As functional timing side channels
are hard to detect statically with the current program analysis
tools, we turn to dynamic analysis methods.
We propose to use gray-box evolutionary search algorithms~\cite{AFL,kersten2017poster}
to generate interesting secret and public inputs.
We use {\em functional data clustering}~\cite{jacques2014functional,ferraty2006nonparametric}
to model functional observations, discover timing side channels, and
estimate their strengths.
It allows us to compute an equivalence relation on secret inputs that model the
distinguishing power of the attacker. If this relation has multiple equivalence
classes, there is an information leak.
In order to find what parts of the code
caused the leak, we identify features that are common for secrets in the same
cluster (equivalence class), and features that separate the clusters.
Typical features in the debugging context are program internals such as
methods called or basic blocks executed for a given secret value.
We present functional extensions to 
{\it decision tree inference techniques} to locate code regions that explain
differences among clusters.
These code regions are thus suspect of being a root cause of the side channels.

\noindent{\bf Experiments.} We evaluate our tool \toolname on
micro-benchmarks and $10$ larger case studies. We use micro-benchmarks to
evaluate the scalability of components in our tool and compare
\toolname to the state-of-the-art.
The case studies serve to evaluate scalability
and usefulness on real-world Java applications.

\begin{figure*}[t!]
  \centering
  \begin{subfigure}{0.25\textwidth}
    \includegraphics[width=0.95\textwidth]{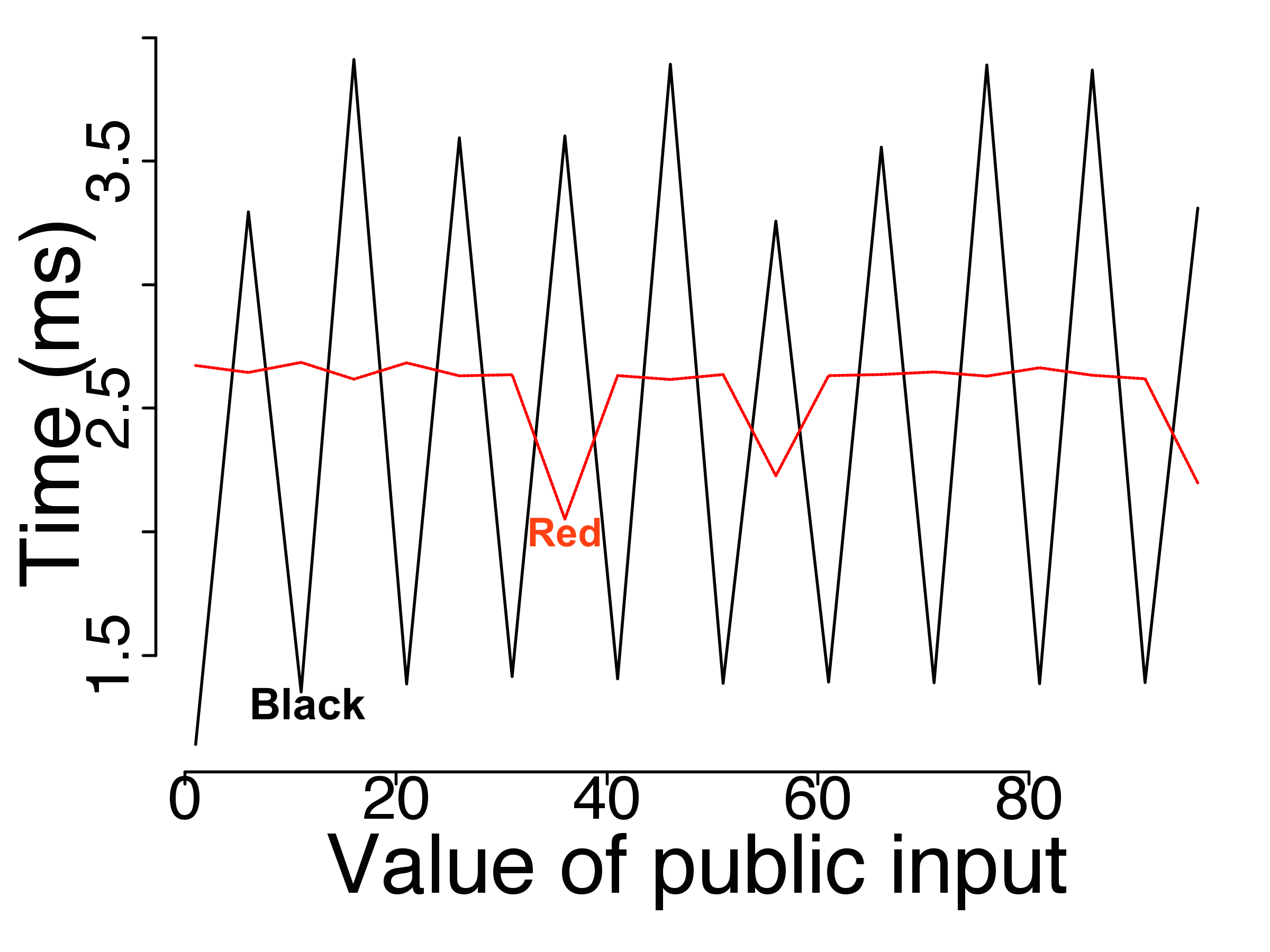}
    \newsubcap{Functional observations.}
    \label{fig:zigzag}
  \end{subfigure}
  \begin{subfigure}{0.25\textwidth}
    \includegraphics[width=0.95\textwidth]{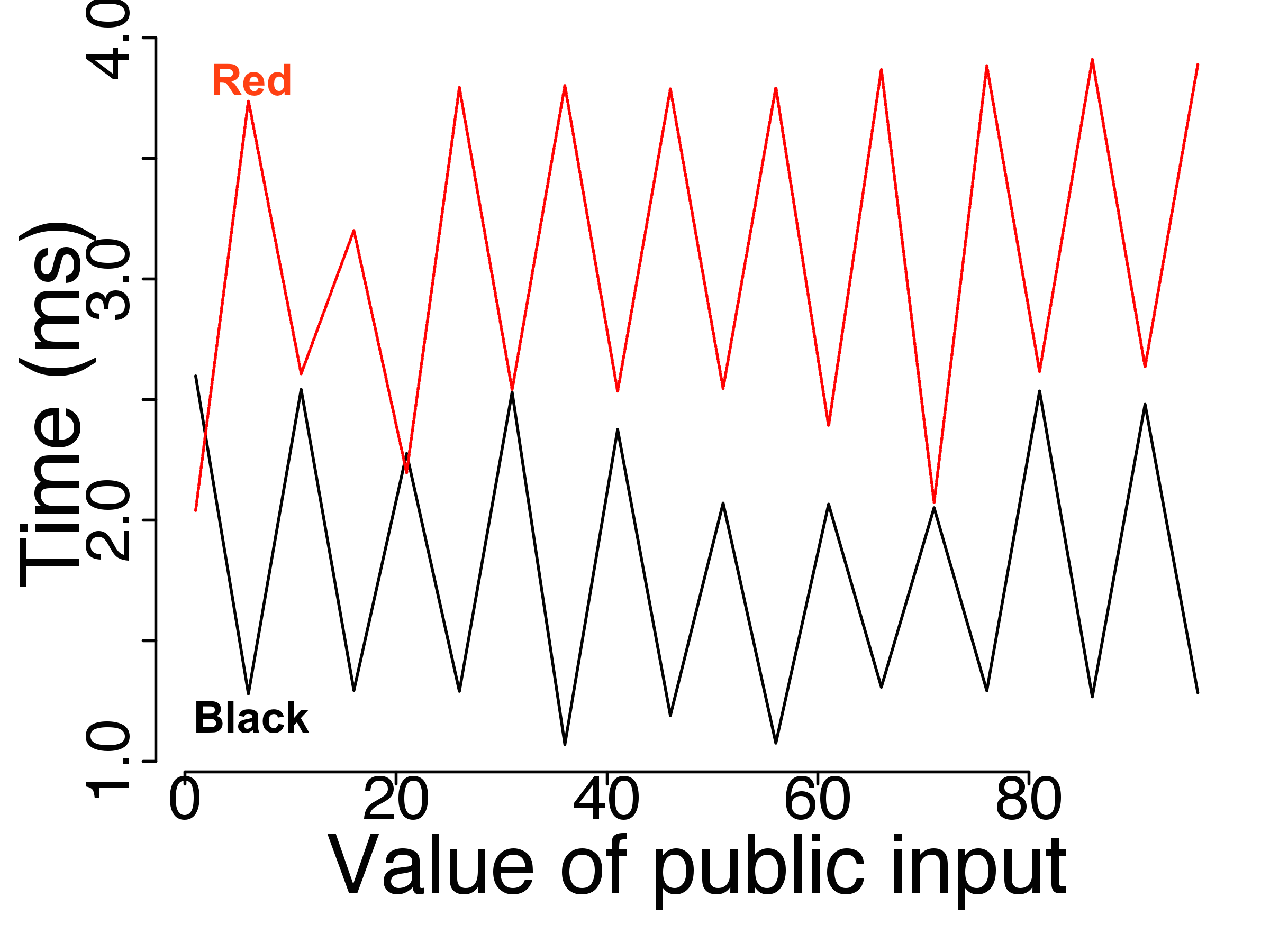}
    \caption{Attacker's local  observations.}
    \label{fig:deriv-example-b}
  \end{subfigure}
  \begin{subfigure}{0.25\textwidth}
    \includegraphics[width=0.95\textwidth]{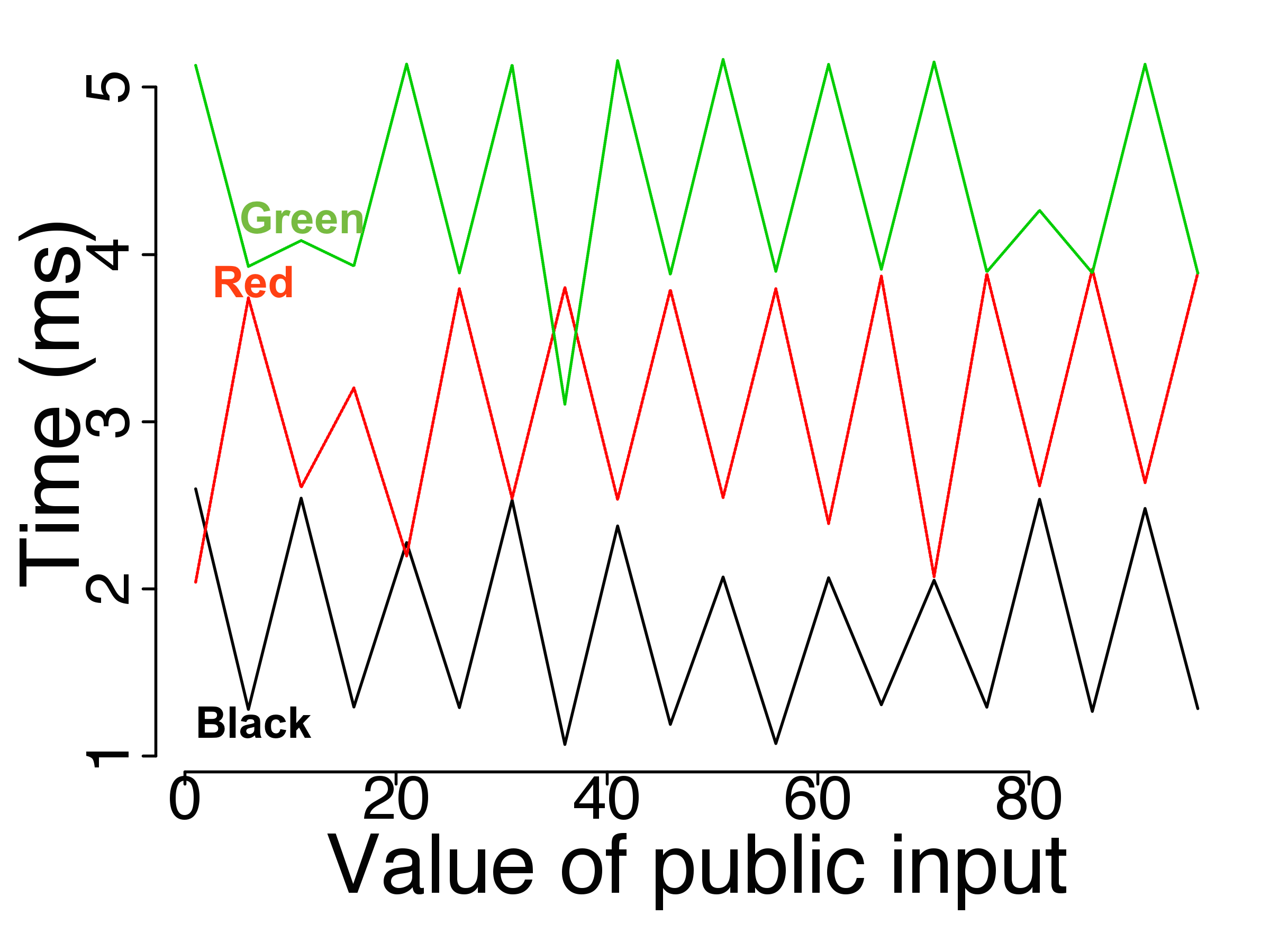}
    \caption{Attacker's remote observations.}
    \label{fig:deriv-example-c}
  \end{subfigure}
 \vspace{-1.0em}
\end{figure*}

\noindent{\bf Contributions.}
Our main contributions are:
\begin{itemize}
	\item Defining {\em functional noninterference} in the presence of noisy
	observation: We demonstrate functional side channels caught by our
	definitions in programs that would be deemed information-leak-free
	or underestimated using the standard (non-functional) definition.
	\item Algorithms: We adapt existing theory and algorithms for {\em functional
	data clustering} to discover the existence of side channels.
	We develop a {\em functional extension of decision tree learning} to locate
	the code regions causing a side channel if there is one.
	\item Experiments: we show on micro-benchmarks that
	\toolname outperforms DifFuzz~\cite{DBLP:conf/icse/nilizadeh},
	a state-of-the-art technique, in quantifying the strength
	of leaks using the number of classes in timing observations.
	\item Case Studies: We show the scalability of \toolname
	in analyzing Java programs with thousands of methods.
	\toolname finds a zero-day vulnerability
	in OpenJDK and vulnerability in Java web-server that was
	since fixed by the original developers.
	\end{itemize}

\section{Functional Side Channels}
\label{sec:motivate}
\begin{figure*}[t!]
  \begin{minipage}{0.82\textwidth}
    \raggedright
    \begin{subfigure}{0.3\textwidth}
      \includegraphics[width=0.9\textwidth]{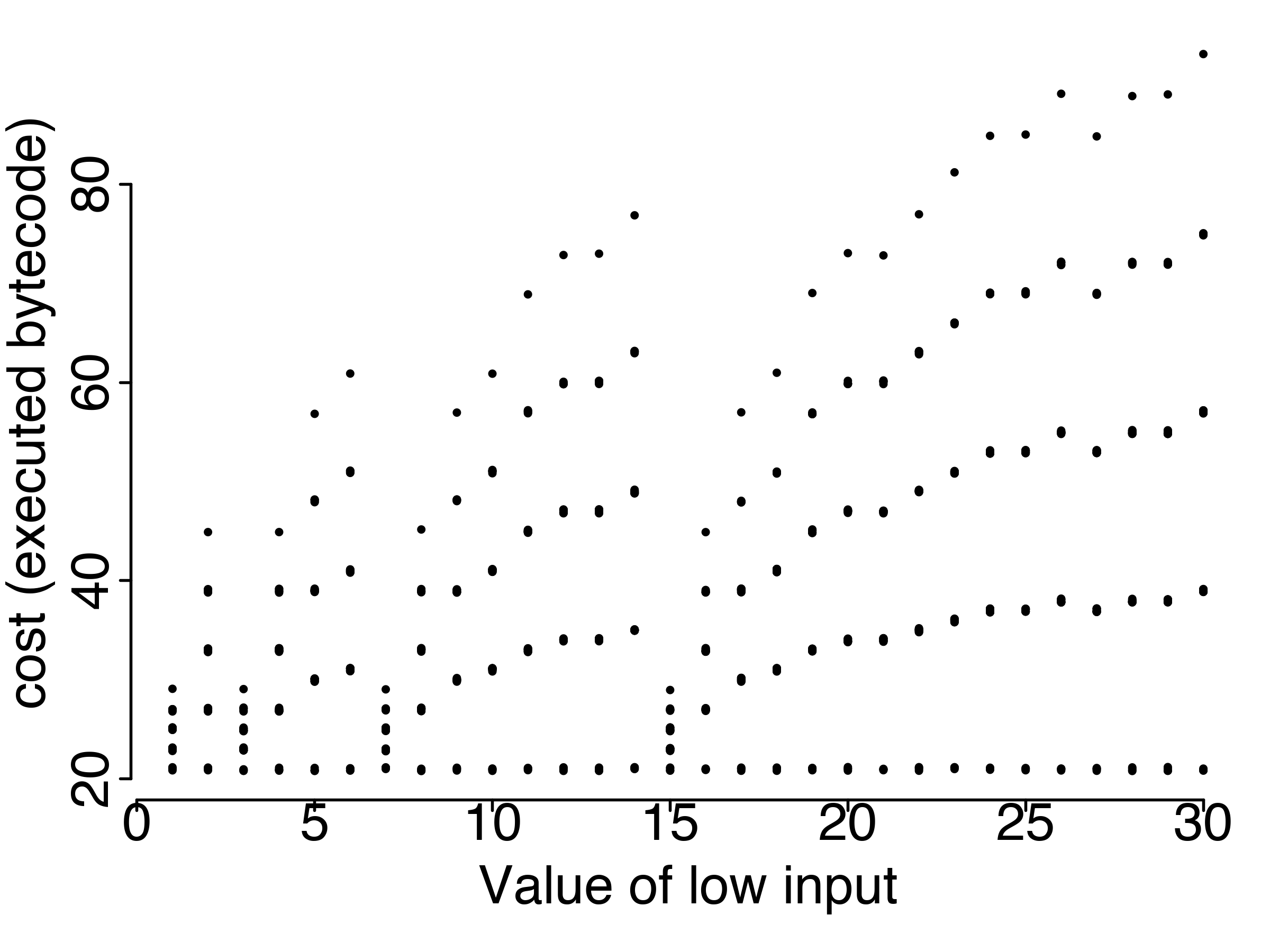}
      \caption{Point-wise model $\Pp_1$.}
      \label{fig:pw-p1}
    \end{subfigure}
    \begin{subfigure}{0.3\textwidth}
      \includegraphics[width=0.9\textwidth]{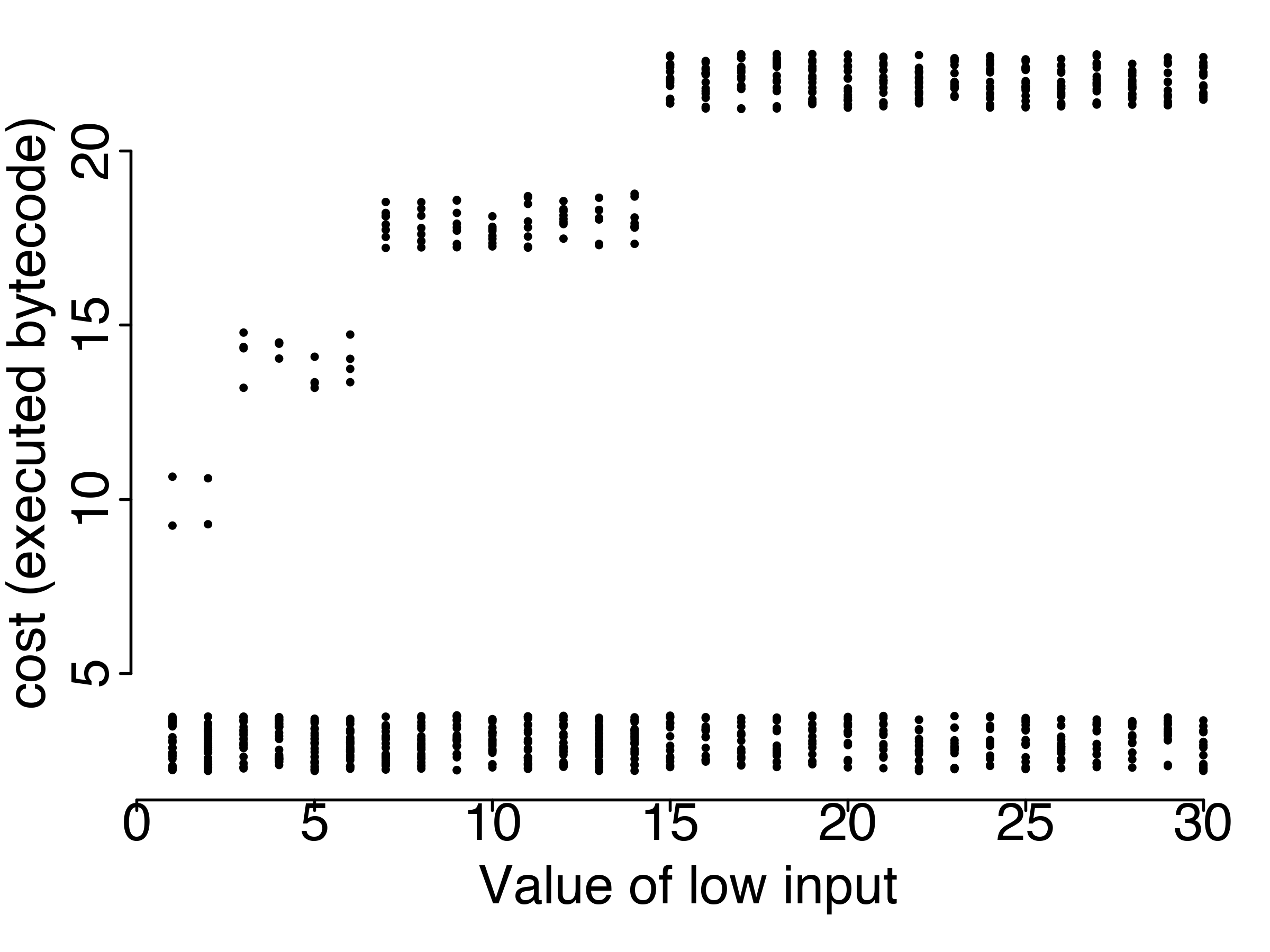}
      \caption{Point-wise model $\Pp_2$.}
      \label{fig:pw-p2}
    \end{subfigure}
    \begin{subfigure}{0.3\textwidth}
      \includegraphics[width=0.9\textwidth]{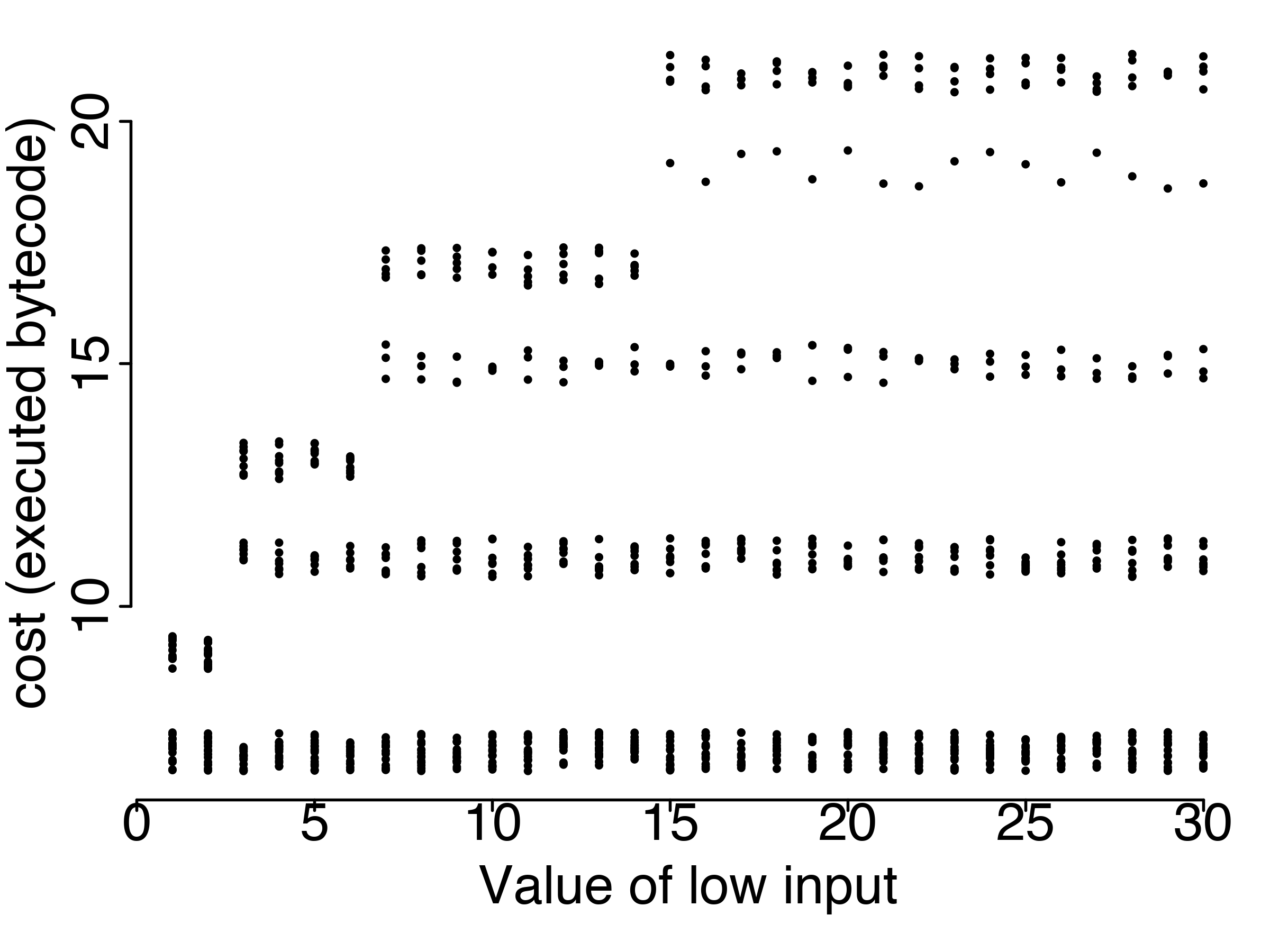}
      \caption{Point-wise model $\Pp_3$.}
      \label{fig:pw-p3}
    \end{subfigure}
    \begin{subfigure}{0.3\textwidth}
      \includegraphics[width=0.9\textwidth]{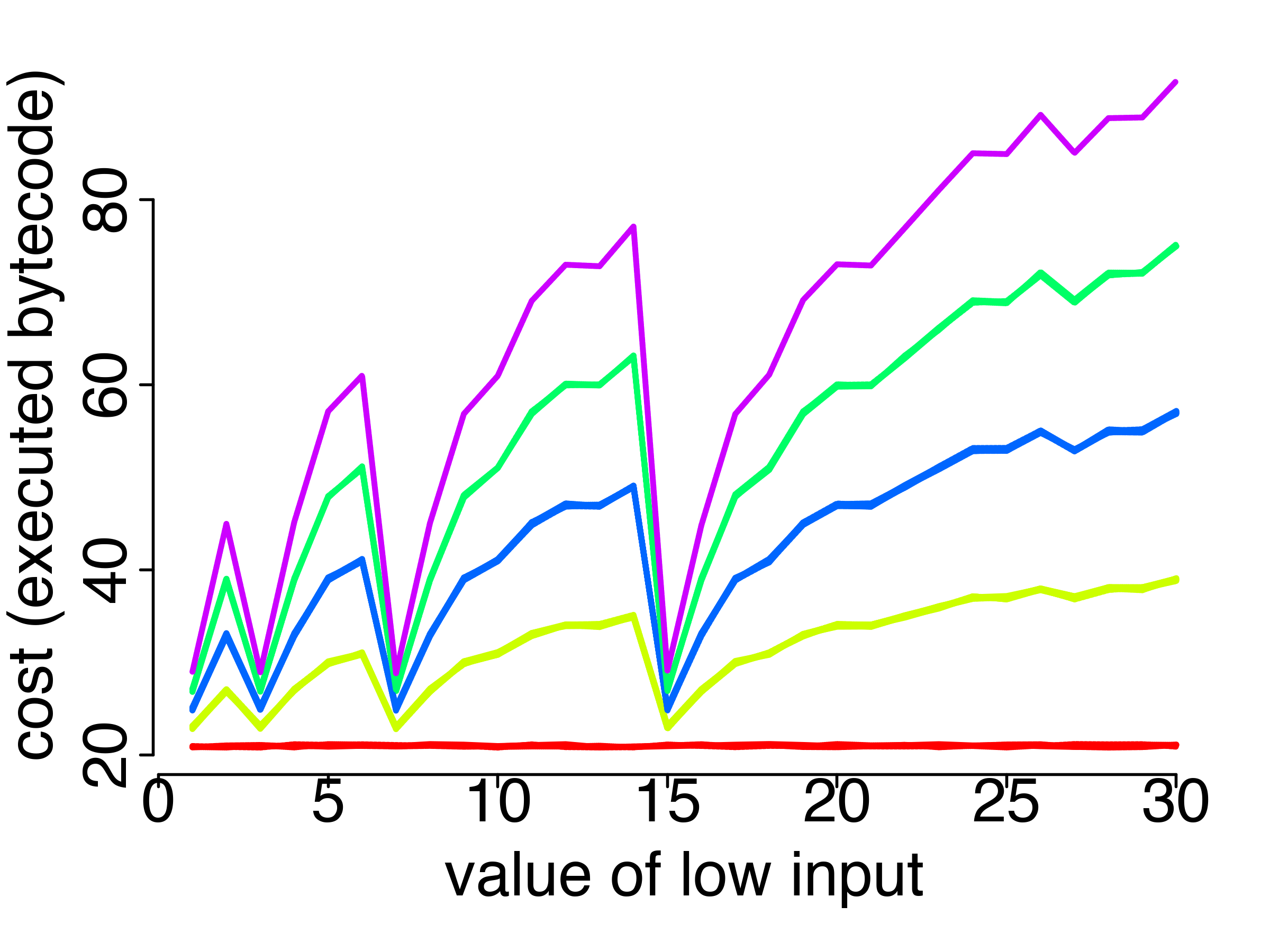}
      \caption{Functional model $\Pp_1$.}
      \label{fig:f-p1}
    \end{subfigure}
    \begin{subfigure}{0.3\textwidth}
      \includegraphics[width=0.9\textwidth]{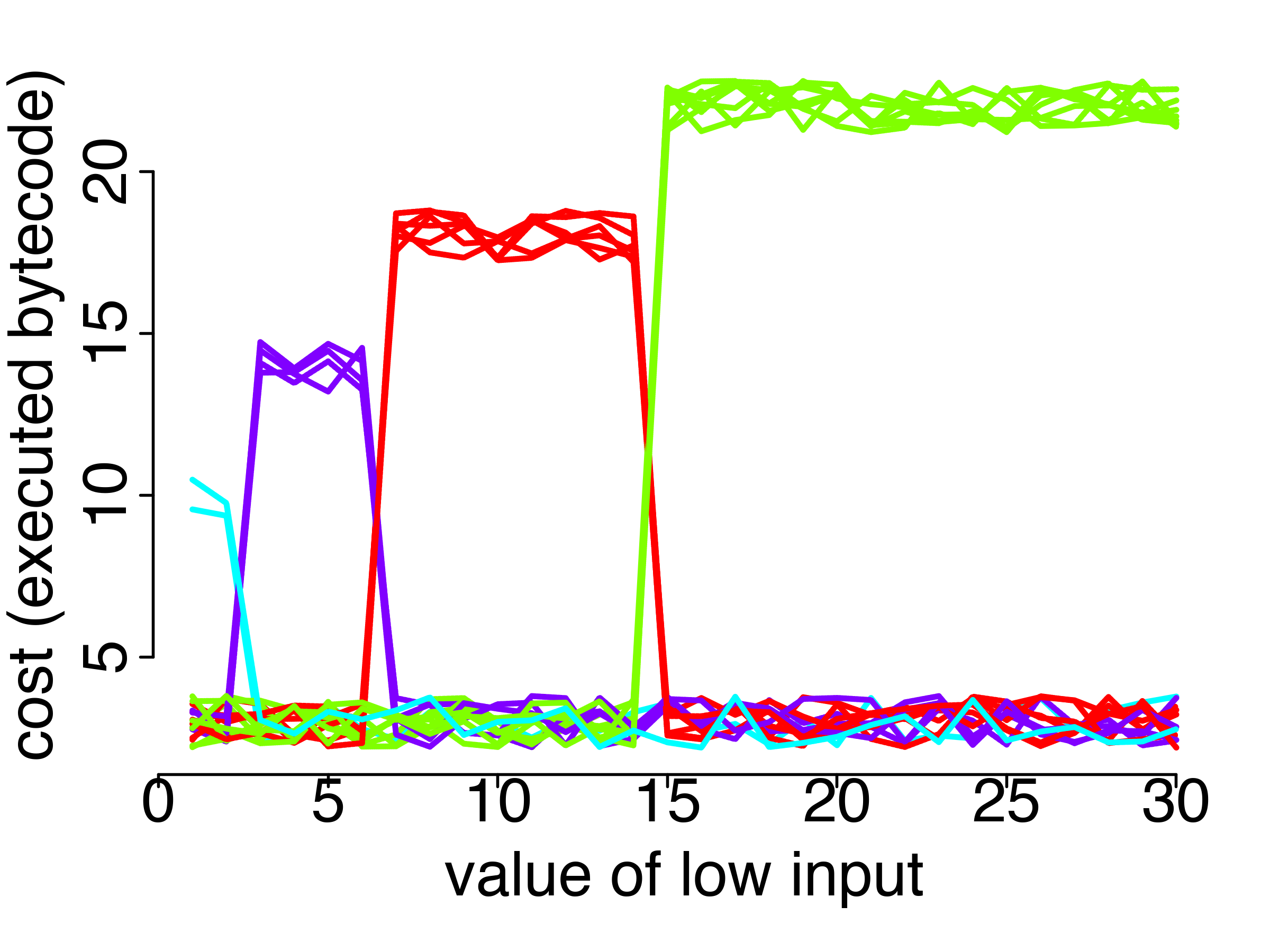}
      \caption{Functional model $\Pp_2$.}
      \label{fig:f-p2}
    \end{subfigure}
    \begin{subfigure}{0.3\textwidth}
      \includegraphics[width=0.95\textwidth]{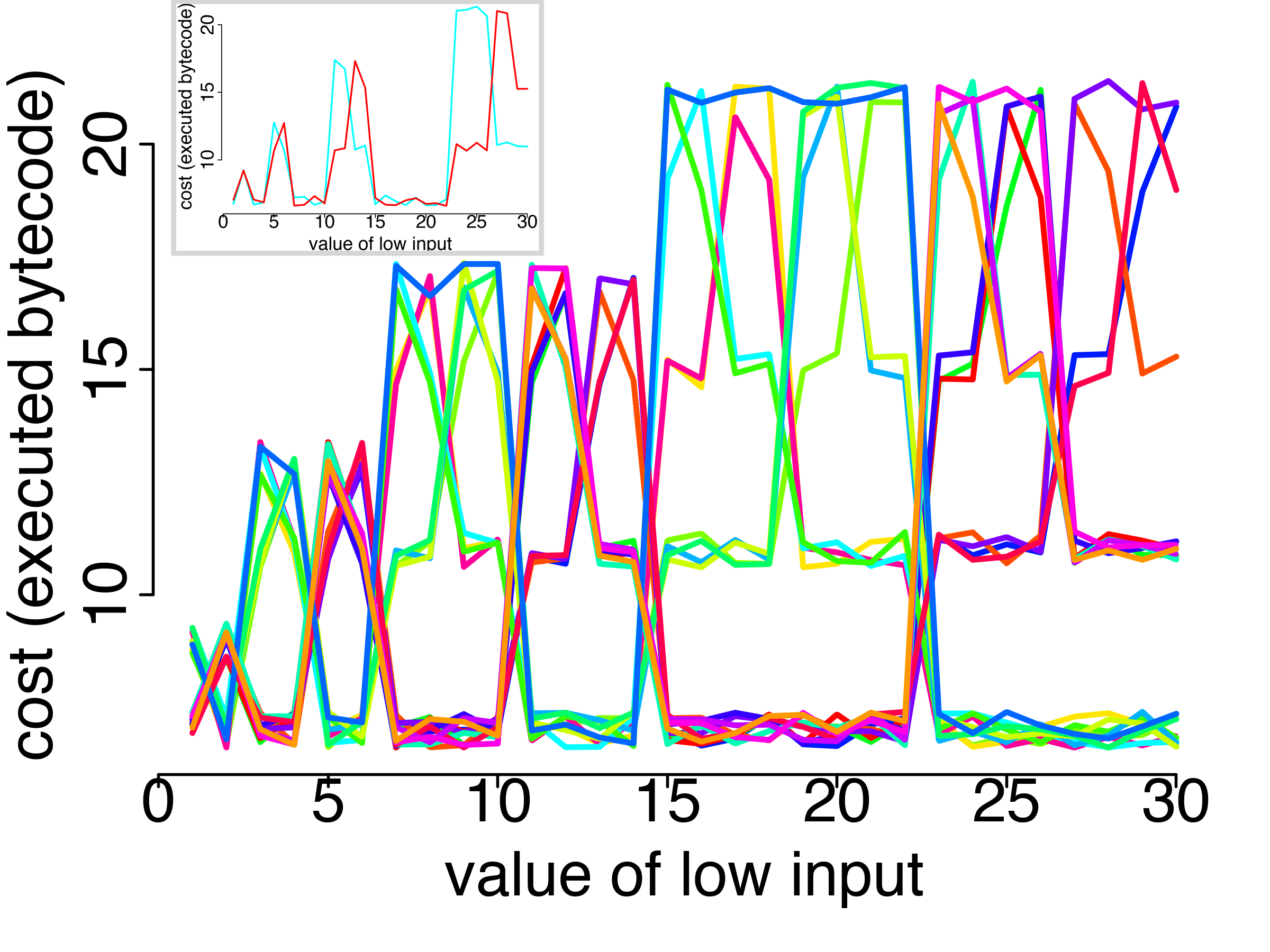}
      \caption{Functional model $\Pp_3$.}
      \label{fig:f-p3}
    \end{subfigure}
  \end{minipage}
  \begin{minipage}{0.17\textwidth}
    \raggedright
    \scalebox{0.8}{
     \begin{tabular}{|c c |}
     \hline
     x-axis & low input value \\ [0.5ex]
     \hline
     1 & ``0'' \\
     \hline
     2 & ``1'' \\
     \hline
     3 & ``00'' \\
     \hline
     4 & ``01'' \\
     \hline
     5 & ``10'' \\
     \hline
     6 & ``11'' \\
     \hline
     7 & ``000'' \\
     \hline
     8 & ``001'' \\
     \hline
     ... & ... \\
     \hline
     14 & ``111'' \\
     \hline
     15 & ``0000'' \\
     \hline
     16 & ``0001'' \\
     \hline
     ... & ... \\
     \hline
     30 & ``1111'' \\
     \hline
    \end{tabular}}
  \end{minipage}
  \vspace{-0.5em}
  \caption{Programs $\Pp_1$, $\Pp_2$, $\Pp_3$ are leaking the number of
            set bits, the length of secret, and the secret values, respectively.
            In programs $\Pp_2$ and $\Pp_3$, the clustering over functions
            are required to find the correct number of classes of observations.}
  \label{fig:illust-clust}
 \vspace{-1.5em}
\end{figure*}

We first illustrate what an attacker can infer based on functional
observations, even in the presence of noise.
Second, we show
that it is critical to use functional observations to evaluate
the resulting threats of side channels.

\subsection{Functional observations and timing side channels}
\label{func-side-channel-epsilon}
We consider the {\em known-message} threats~\cite{kopf2009provably}
where the attacker knows
public values, but she cannot control them.
We focus on the situation where secret values remained
unchanged for some amount of time (e.g., passwords, social security number, and
friends of a user in social media). The attacker who has access to the source code
tries to infer the secret by observing
the execution time and knowing the public values.

Let us consider the classical definition of confidentiality:--
noninterference. A program is {\em unsafe} iff for a pair of secret values
$s_1$ and $s_2$, there exists a public value $p$ such that the behavior of the
program on $(s_1,p)$ is observably different than on $(s_2,p)$.
If our observable is the execution time $T$, then the program is unsafe iff:
$\exists s_1, s_2, p: T(s_1,p) \neq T(s_2,p)$.
In our setting, for each secret value, we
observe the execution time of a program on a number of
public values. Thus, the program is unsafe iff:
%
$\exists s_1, s_2: (\lambda p. T(s_1,p)) \neq (\lambda p. T(s_2,p))$.
%
In other words, the program is unsafe if the two secret values
do not correspond
to the same function of public inputs.

\noindent\textbf{Side channels in the presence of noise.}
Quantitative observations of a program's runtime behavior are often noisy. For
instance, running a program with the same inputs twice on the same machine result in different measurements of execution time.
Observing the program remotely adds a further level of noise.
Classical definitions of confidentiality properties, therefore, need to be adapted
to noisy environments. In the noisy environment, no two observations are equal
and our definition needs to include $\varepsilon$ tolerance:
 \begin{equation}
\exists s_1, s_2: d(\lambda p. T(s_1,p), \lambda p. T(s_2,p)) > \epsilon.
 \label{eq:nonint-func-eps}
 \end{equation}
In this definition, $d$ is a distance between two functions.
The distance is suitably chosen, typically based on the noise
expected for a particular use case.

A straightforward extension of classical noninterference with
$\epsilon$ tolerance to our functional setting is to use the $\infty$-norm for the
distance function:
%
$d_{\infty}(f_{s_1},f_{s_2}) = \sup_p |f_{s_1}(p)-f_{s_2}(p)|$,
where $f_s(p) = \lambda p. T(s,p)$.
However, we now demonstrate that the point-wise
distance $d_{\infty}$ is not the only option, and that depending on the
type of noise, different distances are needed.
%
%
In particular, we show that if we use the point-wise distance, we
could certify a side-channel vulnerable program.

\textit{Gaussian noise (pointwise independent, mean 0)}.
Consider the two functional observations (red and black) of a program in
Figure~\ref{fig:zigzag}. The red function corresponds to the secret value $s_1$
and the black function corresponds to the secret value $s_2$. The eavesdropper can
produce this graph easily by trying possible inputs on their machine beforehand.
At runtime, the eavesdropper collects the public inputs and the
execution time, and tries to learn the secret by matching the observed data to the
red or black functions.
In this example, we assume that the noise for each pair of public-secret inputs
is independent and identically distributed.
Furthermore, we assume that it is distributed according to a Gaussian
distribution with mean $0$.  Let us consider $\varepsilon$ of $3$ms, and then
apply our definition with distance $d_{\infty}$.
We see that the two functional observations are $\varepsilon$-close for
this distance, so the attacker cannot infer the secret value ($s_1$ or $s_2$).
However, the functional observations are clearly very different, and an
eavesdropper can reliably learn the secret.
This can be captured using the $L_1$-norm as
the distance function.
This example shows the point-wise distance $d_{\infty}$
may not be appropriate to detect certain side channels.


\textit{Gaussian noise (pointwise independent, mean C)}.
Let us consider the case where the noise is Gaussian, but with a
non-zero mean. The non-zero mean is fixed, but is
unknown to the attacker. This case arises if, for instance, the eavesdropper is
remote and cannot determine the delay introduced by
the network and separate it from the noise of the remote machine.
 Consider a program with two functional behaviors (red and black)
 pictured in Figure~\ref{fig:deriv-example-b}, where the red and black behaviors
 correspond to secret $s_1$ and secret $s_2$, respectively.

 At runtime, the attacker interacts with the remote server running the same
 instance of the application with a fixed secret value. The green timing
 function in Figure~\ref{fig:deriv-example-c} shows the
 execution-time function of public inputs obtained from observing
 the remote server. The green function looks far apart from both local
 observations (black and red functions).
 However, due to the effects of remote observations, the
 attacker knows that the execution time is off by an unknown
 constant.
 The attacker is in effect observing only the shapes of functions,
 i.e., their first derivatives. So, the distance is over the
 derivatives of functions~\cite{ferraty2006nonparametric,deutilities}.
 The attacker can use this distance and calculate that the green function
 is closer to the black function than the red function. Note that
 if the $L_1$ distance between the first derivative of two
 timing functions is greater than $\varepsilon$,
 the corresponding secret values can leak to a remote attacker.


\subsection{Classes of observations in side channels}
\label{func-side-channel-clusters}
The number of distinct timing observations (or clusters) over secret inputs
is an important measure to evaluate the resulting threats
of side channels~\cite{smith2009foundations,kopf2009provably}.
Here, we illustrate that it is critical to analyze functions to
obtain clusters, especially in dynamic analysis.
We consider three side-channel vulnerable programs.
Program $\Pp_1$, a variant of square and multiply algorithm~\cite{kocher1996timing},
leaks the number of set bits in the secret.
Program $\Pp_2$, a vulnerable Jetty password matcher~\cite{jetty-1},
leaks the length of secret passwords.
Program $\Pp_3$, a vulnerable google Keyczar password
matcher~\cite{Keyczar}, leaks the value of secret passwords.
Let public values be the sequence $\seq{``0",``1",``00",``01",\ldots,``1111"}$,
and secret values be the set $\set{``0",``1",``00",``01",\ldots,``1111"}$.

Figure~\ref{fig:illust-clust}
shows six plots about the execution times of three programs:
Figure~\ref{fig:illust-clust} (a-c) are point-wise depictions
and Figure~\ref{fig:illust-clust} (d-f) are functional presentations.
The x-axis is the index of $30$ public values (see corresponding values
in the table in Figure~\ref{fig:illust-clust}). The y-axis is the cost of executing
a pair of secret and public values in the number of bytecode executed.

According to the point-wise definition, the number of clusters can be
obtained by fixing the public input and finding the number of distinguishable
classes of observations (different cost values) over all possible secrets.
Let's consider a public input value that gives the largest number of clusters
for each example. Let's pick the index $30$ on the x-axis for
all examples in Figure~\ref{fig:illust-clust} (a-c). This choice results in
$5$, $2$, and $5$ clusters for programs $\Pp_1$, $\Pp_2$, and $\Pp_3$,
respectively. According to the functional definition, we model the execution times of each
secret value as a function from the public input to the cost of execution.
Figure~\ref{fig:illust-clust} (d-f) show $30$ functions in each plot, colored
based on their cluster labels. Any two functions that are $\epsilon=0.1$
close to each other belong to the same cluster.

\begin{figure*}[t!]
  \centering
  \begin{subfigure}{0.3\textwidth}
    \includegraphics[width=0.95\textwidth]{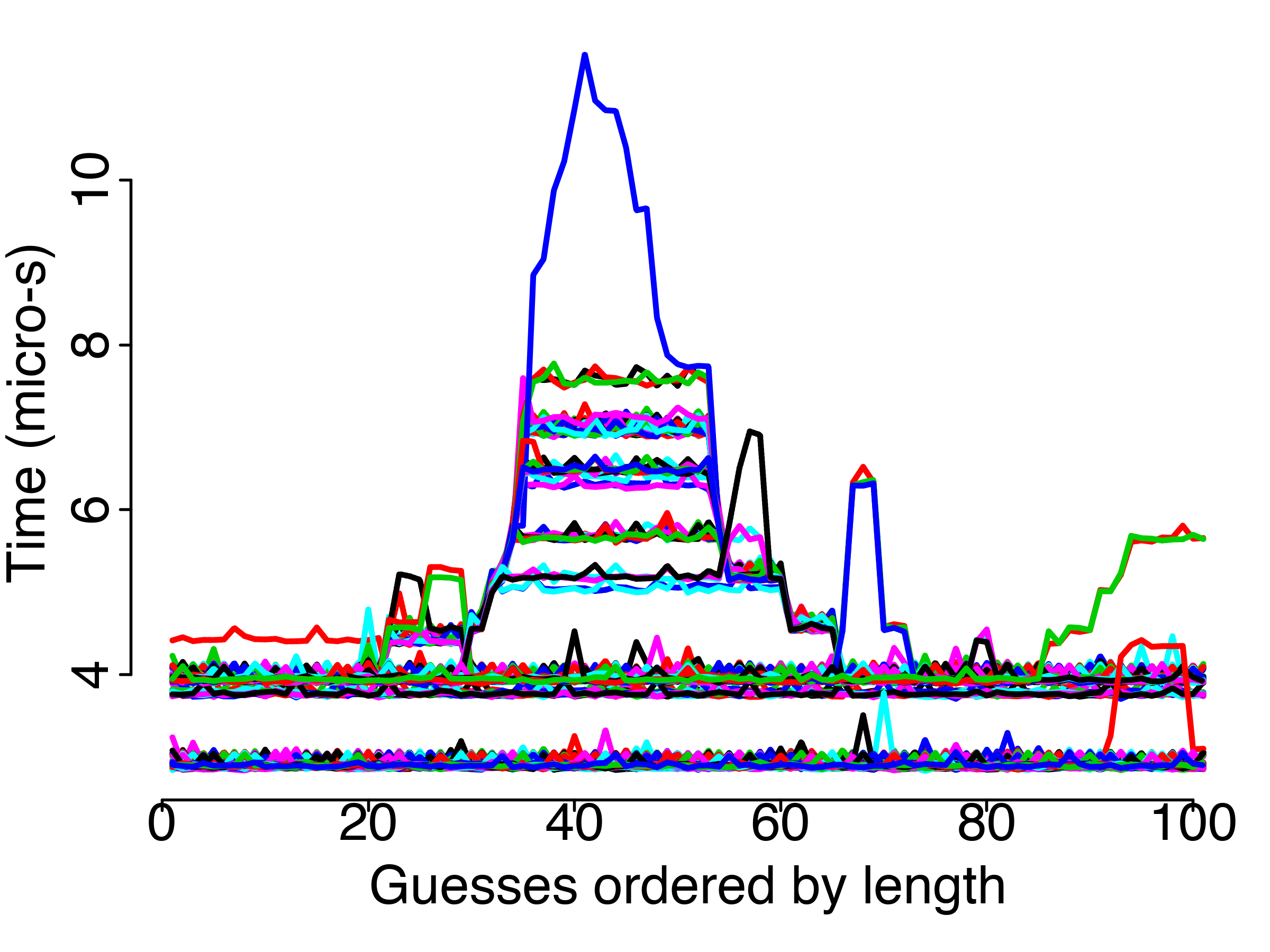}
    \caption{Raw Timing Functions.}
  \end{subfigure}
  \hfill
  \begin{subfigure}{0.3\textwidth}
    \includegraphics[width=0.95\textwidth]{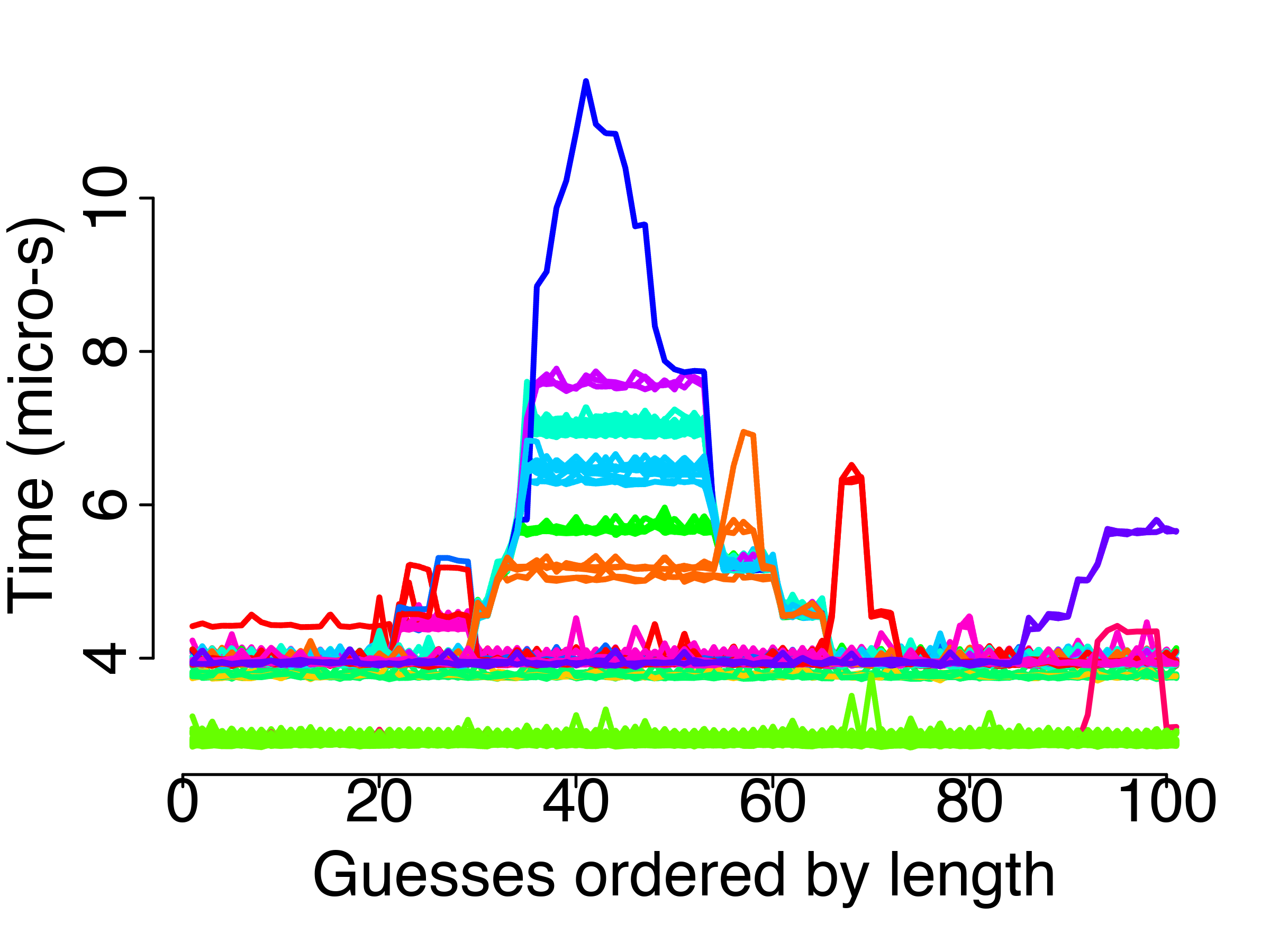}
    \caption{Clustered Timing Functions.}
  \end{subfigure}
  \hfill
  \begin{subfigure}{0.3\textwidth}
  \scalebox{0.55}{
    \begin{tikzpicture}[align=center,node distance=1cm,->,thick,
        draw = black!60, fill=black!60]
      \centering
      \pgfsetarrowsend{latex}
      \pgfsetlinewidth{0.3ex}
      \pgfpathmoveto{\pgfpointorigin}

      \node[dtreenode,initial above,initial text={end parts of tree}] at (0,0) (l0)  {
        java.util.regex.Pattern.\\Slice.match\_bblock\_3964};
      \node[dtreenode,below=of l0] (l2)
       {java.util.regex.Pattern.\\Slice.match\_bblock\_3964};
      \node[dtreenode,below=of l2] (l4)
       {java.util.regex.Pattern.\\Slice.match\_bblock\_3964};
        \node[below=of l2] (l6) {};

        \node[dtreeleaf,bicolor={lightblue and lightblue and 0.99},below left=of l0] (l1) {};
        \node[dtreeleaf,bicolor={aquamarine and aquamarine and 0.99},below left=of l2] (l3) {};
        \node[dtreeleaf,bicolor={violet and violet and 0.99},below left=of l4]
        (l5) {};
        \node[dtreeleaf,bicolor={blue and blue and 0.99},below =of
          l4] (l6) {};

        \path[->]  (l0) edge  node [left,pos=0.4] {$ \leq 19 ~~$} (l1);
        \path  (l0) edge  node [right, pos=0.4] {$~~ > 19 $} (l2);
        \path  (l2) edge  node [left] {$ \leq 22 ~~$} (l3);
        \path  (l2) edge  node [right] {$~~ > 22 $} (l4);
        \path  (l4) edge  node [left] {$ \leq 26 ~~$} (l5);
        \path  (l4) edge  node [right] {$~~ > 26$} (l6);
    \end{tikzpicture}
  }
  \end{subfigure}
  \caption{
    (a) Regex timing functions. Execution times of 435 secret values
    are modeled as 435 timing functions. How are these functions related?
    (b) 435 timing functions are clustered into $12$ distinguishable classes of
    observations (clusters) using $L_1$-norm distances. The presence of different
    clusters indicate some properties of secret patterns are leaking. What
    properties of secret patterns are leaking?
    (c) Regex decision trees. The number of calls to the basic block at
    line 3964 of \texttt{Pattern.match()} method (shown in Figure~\ref{fig:regex-match}) discriminates different
    clusters. The code region shows the value of secret patterns is leaking.}
  \label{fig:regex-clustered}
  \label{fig:regex-remote-1}
	\label{fig:regex-remote-2}
  \label{fig:regex-DT}
  \vspace{-1.0em}
\end{figure*}

For $\Pp_1$, there are $5$ clusters in Figure~\ref{fig:illust-clust} (a),
and also $5$ functional clusters in Figure~\ref{fig:illust-clust} (d).
This means that both the point-wise and functional definitions agree that
there are five clusters ($0$ to $4$ possible set bits).
However, the results are different for programs $\Pp_2$ and $\Pp_3$.
Figure~\ref{fig:illust-clust} (b) shows that in each point on the public value,
there are two classes of observations (either the lengths of secret and
public inputs matched or not).
On the other hand, the functional model in
Figure~\ref{fig:illust-clust} (e) shows that there are $4$
different functional clusters that correspond to four possible lengths in the
secret values. Specifically, let's look at Figure~\ref{fig:illust-clust} (e)
from the indices $7$ to $14$. These are the low (public) values with the length
of three (from ``000'' to ``111'').
Secret values with length three are checked against these public values,
while all other secret values are rejected immediately.
As a result, the timing functions of secret values with the length
three goes up from $2$ to $17$ at index $7$ and goes down from $17$ to $2$
at index of $15$. Observe that the clustering helps to group timing
functions of secret values with length three in the same class.
Similarly, there are unique indices for secrets with
the lengths $1$, $2$, and $4$ where the timing functions jump up/down
and reveal the length of secrets.

Similarly, the point-wise model for the
program $\Pp_3$ in Figure~\ref{fig:illust-clust} (c) underestimates that the
(whole) secret values are leaking (found $5$ clusters).
The functional model in Figure~\ref{fig:illust-clust} (f) finds $30$ clusters:
there is a unique function for each of the $30$ secret values
(two samples are shown on top of Figure~\ref{fig:illust-clust}(f)).
%
%

\section{Overview}
\label{sec:overview}

We illustrate how our tool can be used for discovering and explaining
a zero-day timing side channel. We analyze the
\texttt{java.util.regex}
\footnote{\url{https://docs.oracle.com/javase/8/docs/api/java/util/regex/Pattern.html}}
package of Open JDK 8. The package
includes 620 methods and over 8,000 lines of code.

\noindent \textbf{Problem.}
The secret input is the regular expression compiled as a pattern,
and the public input is the guess that matches against
the pattern. The attacker's goal is to infer a (fixed) secret (or its
properties) by observing the response time for multiple public inputs.
The key problem is to help the \textit{defender}
discover the existence of such side channels.

\noindent \textbf{Side channel discovery.}
The defender starts choosing a
finite set of secret and public values.
In this example, the defender uses \toolname to generate the set
of secret and public input values. The fuzzing of \toolname
is an extension of AFL~\cite{AFL} and Kelinci~\cite{kersten2017poster}
where it generates multiple public input values for each
secret value to discover the functional
dependencies of response time over public inputs.
The defender obtains 1,154 different secret patterns and 6,365 different matching
guesses during 4 hours of fuzzing.
The lengths of guesses are at most 52 characters.
For each secret value,
\toolname varies 6,365 different guesses and measures the execution time of
regex. Then, it fits B-spline~\cite{ramsay2009functional}
to model the timing functions for each secret.
Figure~\ref{fig:regex-clustered} (a) shows 435
different timing functions
over the public inputs from the guesses with the prefix ``bb..." to ``gg...".
We choose the subsets of secret and public values for simpler presentations.
Next, the defender wants to know how
these timing functions are related and if there are timing side channels.

The defender provides the notion of a distance and the tolerance $\varepsilon$.
In this case, he considers $L_{1}$-norm distance between functions and
the tolerance $\varepsilon = 0.2$. Given the distance norm and the tolerance,
\toolname uses a non-parametric functional clustering
algorithm~\cite{jacques2014functional} to discover classes of observations.
The clustering algorithm finds $162$ clusters over the $1,154$ secret values.
The existence of $162$ distinct classes of observations indicates
the presence of a functional side channel in the regex package.
Figure~\ref{fig:regex-clustered} (b) shows $12$ clusters for the
subset of 435 functions presented in Figure~\ref{fig:regex-clustered} (a).

\begin{figure}[b]
\vspace{-1.0em}
  \raggedright
  \begin{lrbox}{\mybox}%
    \begin{scriptsize}
      \begin{mylisting}[hbox,enhanced,drop shadow]{Regex.Pattern.Slice}
boolean ?{\textbf{match}}?(Matcher matcher, int i, ...
  CharSequence seq){
  ?\indentrule?  int[] buf = buffer;
  ?\indentrule?  int len = buf.length;
  ?\indentrule?  for (int j=0; j<len; j++) {   (line.3964)
  ?\indentrule?    if ((i+j) >= matcher.to){
  ?\indentrule?      matcher.hitEnd = true;
  ?\indentrule?      return false;
  ?\indentrule?    }
  ?\indentrule?    if (buf[j] != seq.charAt(i+j))
  ?\indentrule?      return false;
  ?\indentrule?  }
  ?\indentrule?  return next.match(matcher, i+len, seq);
  }
    \end{mylisting}
    \end{scriptsize}
  \end{lrbox}%
  \scalebox{0.82}{\usebox{\mybox}}
  \caption{Pattern matching using regex
  ({\tt buf} secret, {\tt seq} public).}
  \label{fig:regex-match}
\end{figure}

\noindent \textbf{Side channel explanation.}
Now, the defender wants to know what properties
of secret are leaking through the timing side channels.
The task is to learn the discriminant~\cite{tizpaz2017discriminating}.
It helps the defender localize the code regions
causing to observe different clusters and use the information to
establish facts about the leaks. Specifically,
it shows which features are common for secrets in the same
cluster and which features separate different clusters.
\toolname uses program internal features such as
methods called or basic blocks invoked. These are obtained
by running the same set of secret and public inputs
on the instrumented regex (using Javassist~\cite{chiba2000load}).
The instrumentation provides 203 features about the
internals of regex.
\toolname applies an extension of the decision tree learning algorithm
from~\cite{aaai18}. It produces a decision tree whose nodes are labeled by
program internal features and whose leaves represent
sets of secret values inside a cluster.

Figure~\ref{fig:regex-DT} (c)
shows (parts of) the decision tree model learned for regex.
Using this model, the defender
realizes that the executions of the basic block at line
3964 of {\tt Pattern} class in {\tt Slice.match()} method is what
distinguishes the clusters.
This basic block represents the loop body of the {\tt for}
statement in the method shown in
Figure~\ref{fig:regex-match}. For instance, the
purple cluster, the functions with the second highest execution times
around the index 40 of Figure~\ref{fig:regex-DT} (b), corresponds to
the case where {\tt match\_bblock\_3964} calls
more than 22 times but less than 26 times.
Note that the edge values in the
decision tree are also B-spline functions, but we
map them to their max values for the illustration.
Inspecting the relevant code, the defender
realizes that the matching prefix of secret patterns
is leaking through the side channels.
Hence, an attacker can obtain the (whole) secret patterns,
one part in each step of observations. We reported this vulnerability,
and the OpenJDK security team has confirmed it. Since fixing this
vulnerability requires substantial modifications of library,
the developers suggested to add artificial extra delays when the stored
pattern is secret in order to mitigate it.

\section{Definitions}
\label{sec:statement}
We develop a framework for detecting and explaining information
leaks due to
{\it functional timing observations}.

\subsection{Threat Model}
\label{sec:threat}
We consider the known-message threat~\cite{kopf2009provably}
and assume that the secret inputs are less volatile
than public inputs. Thus, the attacker's observations are functional
where for each secret value, she learns a function from the public
inputs to the execution times.
In our threat model, the attacker, who has access to the source code, can sample
execution times arbitrarily many times on her local
machine with different combinations of secret and public values.
She can thus infer an arbitrarily accurate model of the application's
execution times. During the observations on the target machine,
the attacker intends to guess a fixed secret by observing
the application on multiple public inputs.
These observations may not correspond to her local observations
due to several factors, such as, i) target's machine noises,
ii) network delays, and iii) masking delays added
to every response time to mitigate side channels.

\subsection{Timing Model and Functional Observations}
Let $\Real$ and $\Rplus$ be the set of reals and positive reals.
Variables with unspecified types are assumed to be real-valued.
\begin{definition}
  The {\it timing model } $\sPp$ of a program $\Pp$ is a tuple $(X,
  Y, \Ss, \delta)$ where:
  \begin{itemize}
  \item
    $X {=} \set{x_1, \ldots, x_n}$ is the set of {\it secret-input} variables,
  \item
    $Y = \set{y_1, \ldots, y_m}$ is the set of {\it public-input}
    variables,
  \item
    $\Ss \subseteq \Real^n$ is a finite set of {\it secret-inputs}, and
  \item
    $\delta: \Real^n \times \Real^m \to \Rplus$ is the
    execution-time function of the program as a function of secret and public inputs.
  \end{itemize}
\end{definition}

A {\it functional observation} of the program $\Pp$ for a secret
input $s \in \Ss$ is the function  $\delta(s)$ defined as
$\vy \in \Real^m \mapsto \delta(s, \vy)$.
Let $\Ff$ be the set of all functional observations.
To characterize indistinguishability between two functional
observations, we introduce a (normalized) distance function
$\vd{i}{p}: \Ff \times \Ff \to \Rplus$ on functional observations, for $i,
p \in \Nat$, defined as:
\[
\vd{i}{p} \rmdef (f, g) \mapsto \left(\frac{1}{|Y|}\int_{y \in Y} \left(f^{(i)}(y) - g^{(i)}(y)\right)^p dy \right)^\frac{1}{p},
\vspace{-0.5em}
\]
where $f^{(i)}$ represents $i$-th derivative (wrt $y$) of the function $f$
($0$-th derivative is the function itself) and $|Y|$ is a measure for the domain
of public inputs.
The distance function $\vd{i}{p}$ corresponds to the $p$-norm distance between
$i$-th derivatives of the functional observations.
Given the tolerance $\varepsilon > 0$ and the distance function
parameterized with $i,p \in \Nat$, we say that secrets $s$ and $s'$ are
indistinguishable if $\vd{i}{p}(\delta(s),  \delta(s')) \leq \varepsilon$.

Depending upon the context, as we argued in Section~\ref{func-side-channel-epsilon},
different distance functions may be of interest.
For instance, the distance between first derivatives may be applicable when the
shape of the functional observation is leaking information about the secret
and the second derivatives may be applicable when the number of growth spurts in the
observations is leaking information.
Similarly, in situations where the attacker knows the mitigation model---say,
temporal noises added to the signal are $n$-th order polynomials of the public
inputs---two functional observations whose $n$-th derivatives are close in the
$p$-norm sense may be indistinguishable to the attacker.
Finally, depending upon the specific situation, an analyst may wish to use a
more nuanced notion of distance by taking a weighted combination~\cite{GL14} of
various distance functions characterized by $\vd{i}{p}$.  To keep the technical discourse simple,
we will not introduce such weighted combinations.

\subsection{Noninterference and Functional Observations}
Noninterference is a well-established~\cite{goguen1982security,sabelfeld2003language,terauchi2005secure}
criterion to guarantee the absence of side channels.
A program $\Pp$ is said to satisfy the {\it noninterference property} if:
$\forall \vy \in \Real^m  \forall s, s' \in \Sigma \text{ we have }$ $\delta(s, \vy) = \delta(s', \vy).$
To account for the measurement noises in the observation of the execution time,
it is prudent (see, e.g.,~\cite{DBLP:conf/ccs/ChenFD17}) to relax the notion of
noninterference from exact equality in timing observations to a parameterized
neighborhood.
For a given $\varepsilon {>} 0$, a program $\Pp$
satisfies \textit{$\varepsilon$-approximate noninterference} if:
\begin{align} \label{Noninterference-3}
\forall \vy \in \Real^m  \forall s, s' \in \Sigma \text{ we have
} |\delta(s, \vy) - \delta(s', \vy)| \leq \varepsilon.
\end{align}

We adapt the notion of $\varepsilon$-approximate noninterference in our setting
of functional observations by generalizing previous notions of noninterference.
We say that a program satisfies {\it functional $\varepsilon$-approximate
noninterference} if
\begin{align} \label{Noninterference-4}
\forall s, s' \in \Sigma \text{ we have
} \vd{i}{p}(\delta(s),  \delta(s')) \leq \varepsilon,
\end{align}
where $\vd_{i,p}$ is a distance function over functional observations defined
earlier. For example, the distance $d_{0,\infty}$ in the definition~(\ref{Noninterference-4})
recovers the definition~(\ref{Noninterference-3}).
For the rest of the paper,
we assume a fixed distance function $d$ over functions.

\subsection{Quantifying Information Leakage}
The notion of noninterference requires that the attacker should deduce nothing
about the secret inputs from observing the execution times for various public inputs.
However, one can argue that achieving noninterference is neither possible nor
desirable, because oftentimes, programs need to reveal information that depends
on the secret inputs. We therefore need a notion of information leakage.
The number of distinguishable classes in timing observations often provide
a realistic measure to evaluate the strength of information leaks.
For example, the min-entropy measure~\cite{smith2009foundations} quantifies
the amounts of information leaks based on the number of distinguishable observations.
Our data-driven approach with functional clustering algorithms provides
a lower-bound on the classes of observations.

\section{Data-Driven Discovery and Explanations}
\label{sec:discovery}
The space of program inputs are often too large
(potentially infinite) to exhaustively explore even for medium-sized programs.
This necessitates a data-driven approach for discovery and
explanation of functional side channels.
In the proposed approach, an analyst uses
fuzzing techniques, previously reported issues, or domain knowledge
to obtain a set of secret and public inputs.
In particular, an extension of gray-box evolutionary
search algorithms can be used to generate interesting inputs for
functional side channel analysis.
Our technique then exploits functional patterns in the given
inputs and applies functional data clusterings to discover functional
side channels.
To explain the discovered side channels, our tool instruments the
programs to print information about auxiliary features
(e.g., the number of times a method called or
basic block executed) and apply classification inferences
to localize code regions cause the side channel leaks.
To summarize: given a set of program input traces,
the key computational problems are a) to cluster
traces exhibiting distinguishable timing behaviors and b) to explain
these differences by exploiting richer information based on the auxiliary features.

\textit{Hyper-trace Learning.}
Let $Z{=}{\set{z_1, \ldots, z_r}}$ be the set of auxiliary features.
An {\it execution trace} of a program $\Pp$ is a tuple
\[
(\vx, \vy, \vz, t) \in \Real^n \times \Real^m \times \Real^r \times \Real,
\]
wherein $\vx \in \Sigma \subset \Real^n$ is a value to the secret inputs,
$\vy \in \Real^m$ is a value to the public inputs,
$\vz \in \Real^r$ are the valuations to the auxiliary
features, and $t \in \Rplus$ is the execution time.
We assume the valuations of the auxiliary features deterministically
depend only on the secret and public inputs.
To keep the execution time unaffected from the instrumentations, we
estimate the execution time on the un-instrumented programs.
Let $\Tt$ be a set of execution traces.

As our main objective is to explain the differences on functional observations
due to differences on secret and auxiliary features, we rearrange the
raw execution traces $\Tt$ to functional traces $\Hh$ by combining traces with
common values of the secret inputs.
Functional traces $\Hh$ are hyper-traces---as they summarize multiple program
executions---that model auxiliary and timing values as a function of public inputs.
A {\it hyper-trace} $\tau$ is a tuple
\[
\big(\vx,{(\Aa_i(\vx))_{i=1}^r},{f_T(\vx)}\big)
\in \Real^n \times ([\Real^m \to \Real])^r \times [\Real^m \to \Real],
\]
wherein $\vx$ is a value to
the secret input, $\Aa_i$ and $f_T$ are functions modeling values of auxiliary
features and execution time, respectively, as a function of public inputs for
secret $\vx$.
Computation of hyper-traces from a set of raw-traces is achieved by turning the
discrete vectors of observations (for auxiliary variables as well as execution
time) into smooth functions represented as linear combinations of appropriate
basis functions (e.g. B-spline basis system, Fourier basis functions, and
polynomial bases)~\cite{ramsay2006functional}.
We primarily use B-splines.

\textit{Side-Channel Discovery.}
Given a set $\Hh$ of hyper-traces,
$\Hh = \{(\vx_j,(\Aa_i(\vx_j))_{i=1}^r,f_T(\vx_j))\}_{j=1}^N$,
we use functional data clustering over
$C = \{f_T(\vx_j)\}_{j=1}^N$ to detect different classes of
observations such that hyper-traces within a cluster are $\epsilon$-close
according to the distance $\vd{i}{p}$.

Functional clustering approaches~\cite{jacques2014functional} can be broadly classified
into non-parametric and model-based approaches.
Our tool uses a non-parametric functional clustering and implements two algorithms
to cluster indistinguishable observations.
These algorithms---described in Section~\ref{sec:implement}---take the timing
observations set $C$, an upper bound $K$ on the number of clusters, a distance
function $\vd{i}{p}$, and the tolerance $\varepsilon > 0$ as inputs,
and returns the ``centroids'' of observational functions $\Ff = \set{f_1, f_2,
  \ldots, f_k}$ for $k \leq K$.
Our algorithm guarantees that each centroid $f_{\kappa} \in \Ff$ ($1 \leq \kappa \leq k$)
represents the timing functions for the set of secret values $\Sigma_{\kappa}$ such that $\vx, \vx' \in
\Sigma_{\kappa}$ if and only if $\vd{i}{p}(f_T(\vx),f_T(\vx')) \leq \varepsilon$.

\textit{Side-Channel Explanation.}
A \emph{(hyper) trace discriminant} is defined as a disjoint partitioning of the
auxiliary variables (functional) spaces along with
a functional observation for each partition.
Formally, a trace discriminant $\Psi = (\Ff, \Phi)$ is a set of functional
observations $\Ff = \set{f_1, f_2, \ldots, f_k}$---where each $f_{\kappa}
: \Real^m \to \Rplus$ models the execution time as a
function of the public input---and a partition
$\Phi = \seq{\phi_1, \phi_2, \ldots, \phi_k}$ where each
\[
\phi_{\kappa}\::\: [\Real^m \to \Real]^r \to
\set{\texttt{True}, \texttt{False}}
\]
is a predicate over the functional auxiliary features.
We define $\texttt{size}(\Psi)$ as the number of functions in the
discriminant $\Psi$, i.e. $\texttt{size}(\Psi) = |\Ff| = k$.

Given a hyper-trace $\tau {=} (x,(\Aa_i)_{i=1}^r,f_T)$
and discriminant $\Psi {=} (\Ff, \Phi)$, we define the prediction error
$e(\tau, \Psi)$ as $d_{0,2}(f_T, f_{\kappa})$ where $1 {\leq} \kappa {\leq} k$ is the
index of the unique value in $\Psi$ such that
$(x, (\Aa_i)_{i=1}^r) \models \phi_{\kappa}$ i.e. the predicate
$\phi_{\kappa}$ evaluates to
true for the valuation of secret value $x$ and the functional auxiliary
features $(\Aa_i)_{i=1}^r$. This evaluation triggers the functional observation
$f_{\kappa}$. Given a set of hyper-traces $\Hh = \set{\tau(\vx_j)}_{j=1}^N$, and a
discriminant $\Psi$, we define the fitness of the discriminant as the mean of
prediction errors:
\[
\mu(\Hh, \Psi) = \frac{1}{N} \sum_{i=1}^{N} e(\tau(\vx_j), \Psi).
\]

 \begin{algorithm*}[t!]\normalsize
   {
     \DontPrintSemicolon
     \KwIn{Program $\Pp$, the instrumentation $\Pp'$,
      order of public values $\prec_Y$, cluster bound $K$,
      distance $\vd{i}{p}$, and tolerance $\varepsilon$.
     }
     \KwOut{Discovered timing observations as functional clusters
           and their explanations as decision tree models.}

     $\Pi,\Ss = \textsc{FuncFuzz}(\Pp,\prec_Y)$
     \Comment{Obtain secret and public sets by fuzzing $\Pp$
     given an order over public input domain $\prec_Y$}.\;

     $F = \textsc{ExecTime}(\Pp,\Pi,\Ss)$
     	\Comment{Obtain timing functions $F = \{f_T(s_i)\}_{i=1}^{n}$ by executing $\Pp$
        on the set $\Pi$ for each secret $s_i$.}\;

     ${\Aa = \textsc{ExecAux}(\Pp', \Pi, \Ss)}$
      \Comment{Obtain feature set $\Aa$ by executing
        $\Pp'$ similar to the execution of $\textsc{ExecTime}$.}\;

     $\Ff = \textsc{FDClustering}(T,K,\vd{i}{p},\varepsilon)$
      \Comment{Obtain functional clusters $\Ff = \seq{f_1, f_2, \ldots, f_k}$
        over $F$ given $K$, $\vd{i}{p}$, and $\epsilon$.}\;

     $\Phi = \textsc{DiscLearning}(\Aa,\Ff)$
     \Comment{Obtain discriminant $\Phi = \seq{\phi_1, \phi_2, \ldots, \phi_k}$
        given the set $\Aa$ and functional clusters $\Ff$.}\;

     \Return $\Ff$, $\Phi$\;

     \caption{($\Ff, \Phi$) =  \textsc{\toolname}($\Pp, \Pp', \prec_Y, K,
       \vd{i}{p}, \varepsilon$)}
     \label{alg:overall}
   }
 \end{algorithm*}

\begin{definition}[Discriminant Learning Problem]
Given a set of hyper traces $\Hh$, a bound on the size of the discriminant $K \in \Nat$,
a bound on the error $\delta \in \Real$, the
\emph{discriminant learning problem} is to find a model
$\Psi = (\Ff, \Phi)$ with $\texttt{size}(\Psi) \leq K$  and prediction error
$\mu(\Hh, \Psi) \leq \delta$.
\end{definition}
It follows from Theorem 1 in~\cite{AS14} that the
discriminant learning problem is \textsc{NP-hard}.
For this reason, we propose a practical solution to the discriminant learning
problem by exploiting functional data clustering and decision tree
learning.

For learning the discriminant model, we adapt a decision tree learning
algorithm by converting various functional data-values into categorical
variables.
For the $r$ auxiliary features evaluated for a secret $\vx \in \Sigma$,
$(\vx, (\Aa_i(\vx))_{i=1}^{r})$, our algorithm clusters each auxiliary feature
into $k$ groups by employing functional data clustering~\cite{jacques2014functional}.
Let $(\vx, (L_i(\vx))_{i=1}^{r})$ shows secret value $\vx$ and
categorical feature variable $L_i = \set{\ell^1_i, \ell^2_i, \ldots, \ell^k_i}$
for $i=1,\ldots,r$.
Given the set of traces $(\vx_j, (L_i(\vx_j))_{i=1}^r, f_{\kappa})$
with $r$ categorical auxiliary features and the timing
function labeled with cluster color $\kappa$ ($1 {\leq} \kappa {\leq} k$),
the decision tree inference learns hyper-trace discriminants efficiently.

\noindent\textbf{Overall Algorithm.}
\label{sec:overallAlg}
The workflow of \toolname
is given in Algorithm~\ref{alg:overall}.
We provide a brief overview of each component of \toolname here and
describe the details of implementations in the next section.
Given the program $\Pp$ with the secret and public inputs where
$\prec_Y$ defines an arbitrary order over public input domains,
the procedure $\textsc{FuncFuzz}$ employs a gray-box evolutionary search
algorithm to generate public and secret input values.
The procedure $\textsc{ExecTime}$
models timing functions over the public input set for each secret value
on the program $\Pp$.
The procedure $\textsc{ExecAux}$ produces the auxiliary features (method
calls and basic-block invocations) by executing the same inputs as $\textsc{ExecTime}$
on the instrumented program $\Pp'$. Furthermore,
the procedure $\textsc{ExecAux}$ models the feature evaluations
as functional objects over public inputs.
Given an upper bound on the number of clusters $K$, the
distance function $\vd{i}{p}$, and the tolerance $\epsilon$,
$\textsc{FDClustering}$ applies a functional data clustering algorithm
to find classes of observations
$\Ff = \seq{f_1, f_2, \ldots, f_k}$. Each cluster $f_i$ includes a set of
timing functions (corresponds to a set of secret values).
The procedure $\textsc{DiscLearning}$ learns
a set of discriminant predicates $\seq{\phi_1, \phi_2, \ldots, \phi_k}$,
one predicate for each cluster defined over auxiliary features, using decision tree inferences.

\section{Implementation Details} 
\subsection{Implementations of components in \toolname}
\label{sec:implement}
\noindent\textbf{$\textsc{FuncFuzz}$ component.}
We implement fuzzing for our functional side channel discovery
using an extension of AFL~\cite{AFL} and Kelinci~\cite{kersten2017poster}
similar to DifFuzz~\cite{DBLP:conf/icse/nilizadeh}.
The cost notion is the number of bytecode executed for a given
secret and public pair.
In our fuzzing framework, we generate multiple public values for
each given secret value. Then, we model the cost of each secret
value as a simple linear function (for the efficiency of fuzzer) from the domain of
public values to the cost of execution.
This helps exploit simple functional
dependencies of response time (abstracted in the number of bytecode executed)
on public inputs. During fuzzing, we record the linear
cost functions obtained for different secret values. The fuzzing engine
receives small rewards when the linear model of a secret
value has changed and larger rewards when a new linear model found that
is different than any other models (in the same public input domain) observed so far.
Notice that these rewards are in addition to the internal rewards
in AFL such as when it finds a new path in the program.

\noindent\textbf{$\textsc{ExecTime}$ component.}
For each secret value, we have a vector of execution times over public inputs.
We use functional data analysis tools~\cite{ramsay2009functional} to
create B-spline basis and fit functions to the vector of timing observations.
The bases are a set of linear functions that are
independent of one another. Given a known basis, B-spline models can approximate
any arbitrary functions (see~\cite{ramsay2006functional} for more details).
The output of this step is a set of timing functions each for a
distinct secret.

\noindent\textbf{$\textsc{ExecAux}$ component.}
We use Javassist~\cite{chiba2000load} to instrument any
methods in a given package.
The instrumented program $\Pp'$ provides us
with the feature set $Z$ that is method and basic block calls.
For each secret value, we have a vector of the number of
calls to the basic blocks and methods over the public inputs.
We generally fit B-spline over the
valuations of each auxiliary feature $z \in Z$, but we allow for
simpler functions such as polynomials.
The result of this step is the set $\Aa$ that defines functional
values of auxiliary feature $z \in Z$ in the domain
of public inputs.

\begin{figure*}[!t]
	\centering
	\includegraphics[width=\textwidth]{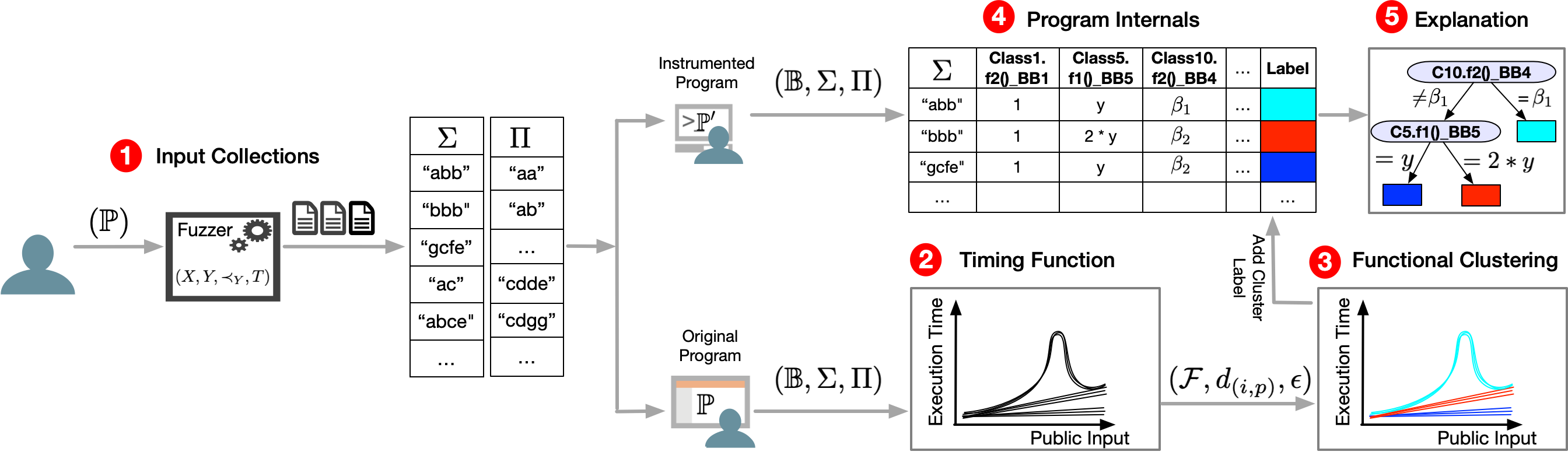}
	\caption{
  \toolname framework. (1) The defender feeds the program $\Pp$ to the fuzzing
  engine and generates a set of secret ($\Sigma$) and public
  ($\Pi$) inputs. (2) The defender specifies the basis function $\mathbb{B}$
  (such as B-Spline basis) and enables \toolname to generate
  timing functions $\Ff$. (3) Given the functions, the defender
  specifies the tolerance $\varepsilon$ and the distance norm
  (such as $L_1$-norm), and \toolname identifies the clusters in timing functions.
  (4) On the instrumented version $\Pp'$, \toolname uses the same
  basis $\mathbb{B}$ and models the calls to each basic block with functions.
  (5) Given the cluster label and the basic block evaluations, \toolname applies
  decision trees to explain side channels with program internal properties.
	}
	\label{fig:FUCHSIA}
\end{figure*}

\noindent\textbf{$\textsc{FDClustering}$ component.}
Given an upper-bound $K$ on the number of clusters and
the distance norm $d$ with the tolerance $\varepsilon$,
we implement $\textsc{FDClustering}$ to discover $k$ clusters ($k \leq K$).
This clustering is an instantiation
of non-parametric functional data clustering~\cite{ferraty2006nonparametric}.
We use two algorithms:
hierarchal~\cite{johnson1967hierarchical}
and constrained K-means~\cite{wagstaff2001constrained}.

\noindent\textit{Preparation for clustering.}
The input for the clustering is the timing functions from $\textsc{ExecTime}$ component.
We use the distance function $\vd{i}{p}$ to obtain the distance matrix $D$.
The distance matrix quantifies the distance between any timing functions.
We specify cannot-link constraints over the matrix $D$.
Cannot-link constraints disallow two functions that
are more than $\varepsilon$ far to be in the same cluster.

\noindent\textit{Constrained K-means clustering.}
Given the upper bound $K$,
constrained K-means algorithm~\cite{song2017package}
obtains $k$ clusters in each iteration ($k$ = 1 in the first iteration).
If the algorithm could not find $k$ clusters ($k \leq K$)
with the cannot-link constraints, it increases $k$ to $k+1$ and
runs the next iteration.
Otherwise, it returns the cluster object  $\Ff = \seq{f_1, f_2, \ldots, f_{k}}$.
The constrained K-means with cannot-link constraints
is known to be computationally intractable~\cite{davidson2005clustering}.

\noindent\textit{Hierarchical clustering.}
The clustering algorithm with complete link method~\cite{R}
obtains $k$ clusters ($k \leq K$).
In each iteration ($k$ = 1 in the first iteration),
it applies the hierarchal clustering, and
then checks the cannot-link constraints to make sure that
all the functions in the cannot-link set are in different clusters.
If the condition is not satisfied,
it increases $k$ to $k+1$ and runs the next iteration.
Otherwise, it returns $\Ff = \seq{f_1, f_2, \ldots, f_{k}}$.
Hierarchical clustering is agnostic to the constraints, and the
constraints are checked after the clustering.

\noindent\textit{Point-wise clustering.}
We use the definition of well-establish $\epsilon$-approximate noninterference
in~\cite{DBLP:conf/ccs/ChenFD17,DBLP:conf/icse/nilizadeh} for point-wise clustering.
For every public input value, we form cannot-link constrains
and apply one of the clustering algorithms
with the $\infty$-norm and the tolerance $\varepsilon$.
Finally, we choose the largest number of clusters among all values
of the public inputs.

\noindent\textbf{$\textsc{DiscLearning}$ component.}
Using the auxiliary variables $\Aa$ as features
and the functional clusters $\Ff$ as labels, the problem of
learning discriminant models becomes a standard classification problem.
The white-box decision tree model explains what auxiliary
features are contributing to
different clusters. We use the CART decision tree algorithm~\cite{Breiman/1984/CART}.

%
%
%
%
%
%
%
%

\noindent\textbf{\toolname framework.} Figure~\ref{fig:FUCHSIA} shows the steps
of \toolname for a defender to discover and explain timing side channels.
\begin{itemize}
	\item[(i)] The user (defender) starts interacting with $\toolname$ by feeding the program or library $\Pp$ to the
  fuzzing engine. This involves modifying the fuzzing driver to call the
  main method in program $\Pp$ with secret input variables $X$, public
  input variables $Y$, and an order on the public input $\prec_Y$. \toolname
  supports all variable types and provides various options for ordering the public
	input variables
	including the size (default), the lexicographic order, and the number of set bits.
	The user can optionally tune the parameter determining the number of public values to be
  generated per each secret value in the fuzzing driver (the default value for this parameter is 3).

  The user then invokes the fuzzer and has an option of stopping it either after a
	pre-specified timeout $T$ (default is 2 hours),
	or when a desired number of inputs is generated. After the fuzzing, the user gathers the set
  of secret ($\Sigma$) and public ($\Pi$) inputs. Optionally, the user
  can specify any other desired set of inputs with unexpected behaviors. 
	\item[(ii)] In the next step, $\toolname$ identifies timing functions for each secret input
	generated in the first step. The defender has the
	option to choose the basis
  $\mathbb{B}$ for timing functions, such as B-splines (default) or polynomials.
	For each secret input,
  \toolname runs the program $\Pp$ on the set of public inputs $\Pi$, measures
  the response times, and fits the basis to generate the timing function.
  The defender obtains the set of timing functions $\Ff$, one for each secret value.
  Optionally, the defender may use the number of byte-code executed instead of
  the actual response time.
	\item[(iii)] Next, \toolname identifies natural clusters in this set of timing
	functions $\Ff$. To aid this, the defender provides the
  distance norm $d_{i,p}$ and a tolerance $\varepsilon$, and \toolname returns
  the cluster label for each timing function. The implemented options
	for the $d_{i,p}$-norm include $L_1$-norm (p = $1$, default), $L_\infty$-norm (p = $\infty$), and $L_2$-norm (p = $2$) for the timing functions ($i=0$, default) and their first derivatives ($i=1$).

	The parameter $\varepsilon$ (with the default value of $0.1$)
	can be fine-tuned based on the
  noises present the timing observations using the following procedure:
  a) select a secret value randomly; b) run the program (with that secret value)
  multiple times on the set of public values; c) create several
  timing functions and employ the clustering algorithm;
  d) search for the smallest value of tolerance $\varepsilon$ such that the
  algorithm returns one cluster. This sampling procedure can be repeated multiple
  times (with different secrets) to get more precise estimates. The accuracy
	of decision tree is another key criterion to base the tuning of the
	$\varepsilon$ parameter and choose values leading to accurate trees.
  \item[(iv)] The fourth step is to generate program internal traces for the inputs reported in the first step. \toolname allows the user to specify {\it features} over program internals---such as basic blocks traversed and set of methods invoked---to base the explanation of the timing side channels. \toolname runs the same set of secret and public inputs on the instrumented program $\Pp'$ to collect data about these features. This results in
  a rich summary of the program traces expressed as the values of these features.
  \toolname uses the same basis $\mathbb{B}$ (default) and model
  the number of calls to each basic block with functions.
  \item[(v)]
	In the last step, given the basic block evaluations and cluster label for each secret value,
  \toolname uses the decision tree models to localize code regions that contribute
  to the creation of timing side channels. This step does not require parameters from the defender.
\end{itemize}
\subsection{Environment Setup}
\label{sec:environment}
All timing measurements in $\textsc{ExecTime}$ of Algorithm~\ref{alg:overall}
are conducted on an Intel NUC5i5RYH~\cite{IntelNUC}.
We run each experiment 10 times and use the mean for the analysis.
All other components are conducted
on an Intel i5-2.7 GHz with 8 GB RAM.
The \toolname includes almost 2,000 lines of code.
The functional analysis and clustering are implemented
in $R$ using functional data analysis package~\cite{fda-usc}
and hierarchal clustering package~\cite{R}.
The fuzzing and instrumentations are implemented
in Java using AFL~\cite{AFL}, Kelinci~\cite{kersten2017poster}, and
Javassist~\cite{chiba1998javassist}.
The decision tree learning algorithm is
implemented in python using scikit-learn library~\cite{scikit-learn}.

\subsection{Micro-benchmarks}
\label{sec:experiment}
\begin{figure}[b!]
\vspace{-1.0em}
	\centering
	\includegraphics[width=0.23\textwidth]{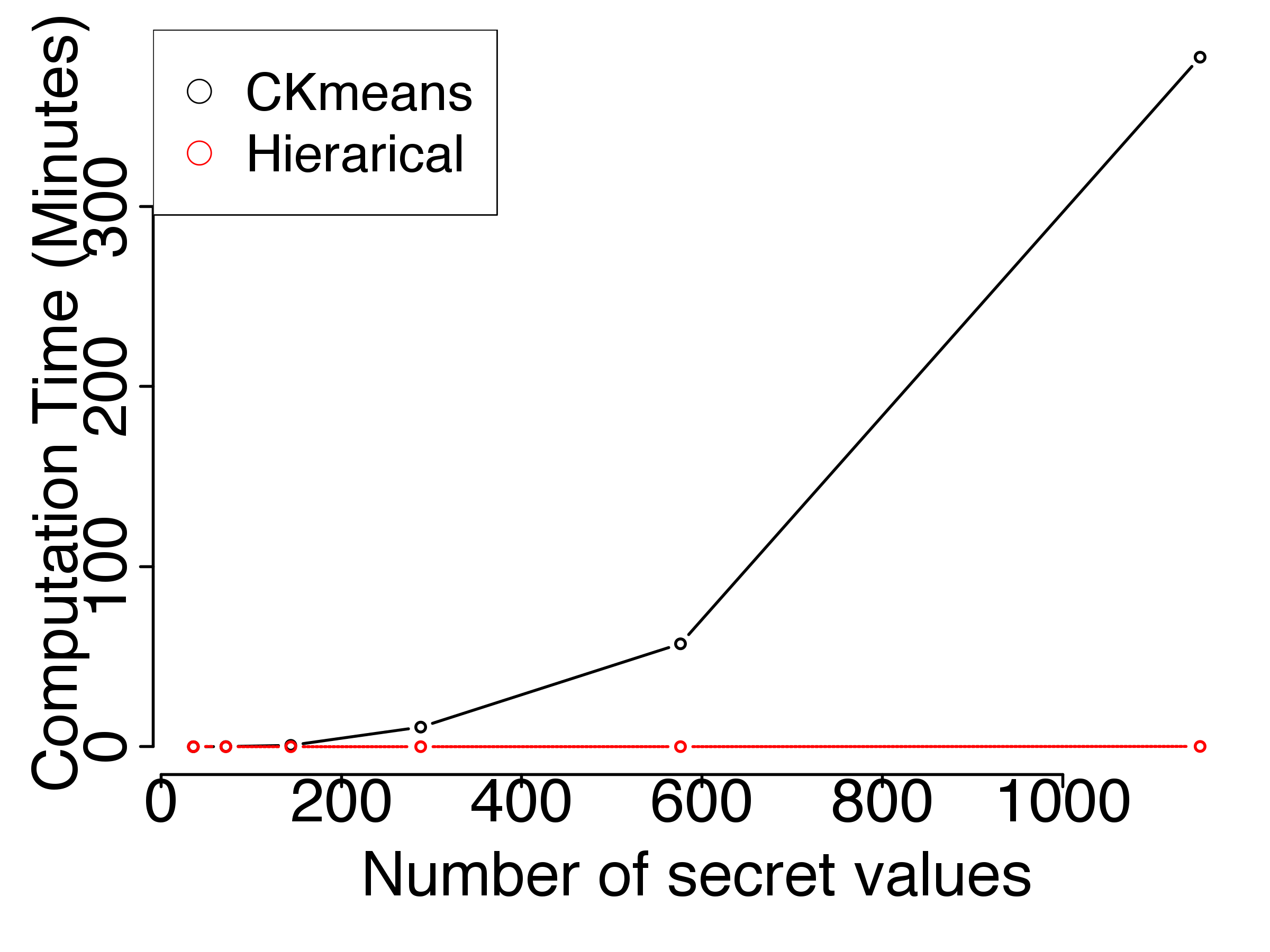}
	\hfill
	\includegraphics[width=0.23\textwidth]{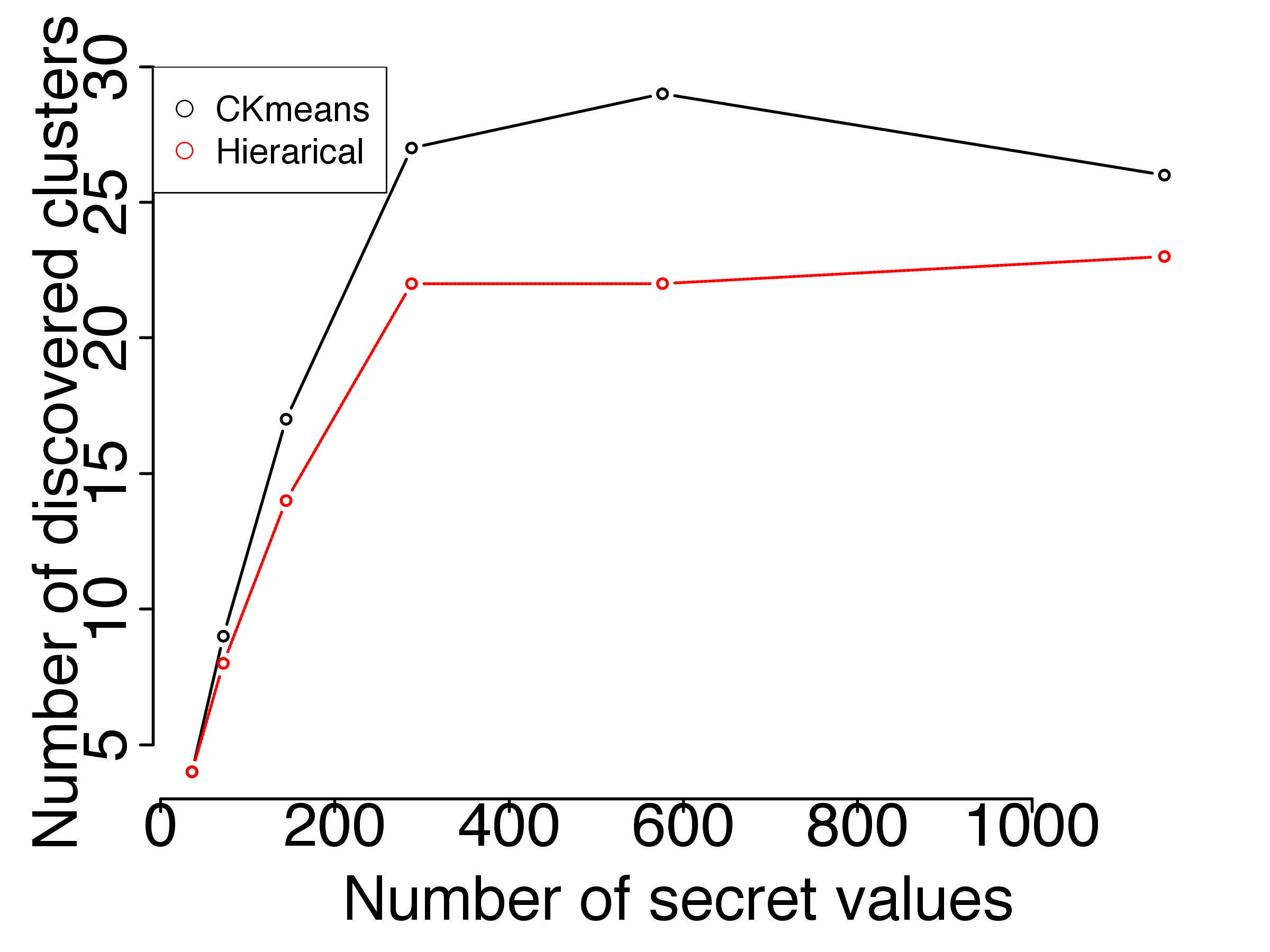}
	\label{fig:comp-kmeans-hierarichal}
	\caption{
		(a) Computation time. Hierarchical clustering is scalable better than
		constrained K-means.
		(b) Number of clusters. Hierarchical clustering discovers a fewer number
		of clusters.
	}
	\label{fig:comp-kmeans-hierarichal}
\end{figure}

We first compare the two clustering algorithms
from Section~\ref{sec:implement}. Then, we examine the
scalability of different components in \toolname. Finally,
we study and compare the results of \toolname versus
DifFuzz~\cite{DBLP:conf/icse/nilizadeh}.

\noindent\textbf{Programs}. Two programs \texttt{Zigzag} and
\texttt{processBid} are shown in Figure~\ref{fig:micro-programs}.
The applications \texttt{Guess\_Secret\_1}~\cite{phan2017synthesis}
and \texttt{Guess\_Secret\_2}~\cite{guess-2}
(shown in Figure~\ref{fig:micro-programs})
take the secret and public as the inputs
and execute different sleep commands depending
on their values.
\texttt{PWCheck\_unsafe} is a password checking example taken
from~\cite{DBLP:conf/icse/nilizadeh}.
Six versions of branch and loop are considered, with
one depicted in Figure~\ref{fig:micro-programs}.
Depending on the secret value, the program does
computations with four types of complexities: O(log(N)), O(N), O(N.log(N)),
and $O(N^2)$ where N is the public input.
Each branch and loop program has all four loop complexities
with different constant factors such as O(log(N)) and O(2.log(N)).

\noindent\textbf{Input Generations}.
To study the micro-benchmark programs, we generate
inputs using our fuzzing technique. We run
the fuzzing for 30 minutes on each program and use the generated
inputs for the rest of analysis.

\noindent\textbf{Clustering Parameters}.
We use both the functional data clusterings (constrained K-means and hierarchical)
as well as point-wise clustering.
For the point-wise clustering, the distance is
based on $\infty$-norm.
For functional clusterings, we consider
the $L_1$-norm distance ($d_{1}$) with the tolerance
$\varepsilon_{1}$.

\begin{figure*}[t!]
\caption{Sample programs used in Micro-benchmark analysis.}
	\centering
        \scalebox{0.63}{\usebox{\mybox}}
        \begin{lrbox}{\mybox}%
	  \begin{scriptsize}
            \begin{mylisting}[hbox,enhanced,drop shadow]{\circled{1}
        Zigzag \hfill \circled{2} processBid}
?{\textbf{Zigzag}}?(int secret, int low) {
  ?\indentrule?if(secret % 2 == 0){
  ?\indentrule?  if (low % 2 == 0) Thread.sleep(3);
  ?\indentrule?  else Thread.sleep(1);}
  ?\indentrule?else Thread.sleep(2);}

?{\textbf{processBid}}?(int sec, int offer){
  ?\indentrule?if (offer < secret) return false;
  ?\indentrule?else { recordBid(offer); return true;}}
           \end{mylisting}
	  \end{scriptsize}
	 \end{lrbox}%
	\scalebox{0.63}{\usebox{\mybox}}
	\begin{lrbox}{\mybox}%
	  \begin{scriptsize}
            \begin{mylisting}[hbox,enhanced,drop shadow]{\circled{3} Branch\_loop\_1}
?{\textbf{Branch\_loop\_1}}?(int secret, int N) {
  ?\indentrule?if (secret<100) for(int i=N; i>0; i /= 2)  Thread.sleep(1);
  ?\indentrule?else if (secret<195) for(int i=0;i<N; i++) Thread.sleep(1);
  ?\indentrule?else if (secret < 290) for (int i = 0; i < N; i++) {
  ?\indentrule?    for (int j = N; j > 0; j /= 2) Thread.sleep(1);
  ?\indentrule?  }
  ?\indentrule?else if (secret < 400) for(int i = 0; i < N; i++) {
  ?\indentrule?    for(int j = 0; j < N; j++) Thread.sleep(1);
  ?\indentrule?}}
              \end{mylisting}
	  \end{scriptsize}
	\end{lrbox}%
        	\scalebox{0.63}{\usebox{\mybox}}
	\begin{lrbox}{\mybox}%
	  \begin{scriptsize}
            \begin{mylisting}[hbox,enhanced,drop shadow]{\circled{4} Guess\_Sec\_2}
?{\textbf{Guess\_Sec\_2}}?(int secret,int low,int t) {
  ?\indentrule?if(low <= secret) {
  ?\indentrule?  if(t == 1) Thread.sleep(1);
  ?\indentrule?  else if (t == 2) Thread.sleep(10);
  ?\indentrule?  else Thread.sleep(1000);}
  ?\indentrule?else {
  ?\indentrule?  if (t == 1) Thread.sleep(1);
  ?\indentrule?  else if(t == 2) Thread.sleep(100);
  ?\indentrule?  else Thread.sleep(1000);}}
              \end{mylisting}
          \end{scriptsize}
	\end{lrbox}%
	\scalebox{0.63}{\usebox{\mybox}}
	\label{fig:micro-programs}
  \vspace{-1.0em}
\end{figure*}

\begin{table*}[t!]
	\centering
		\caption{Micro-benchmark results for \toolname and DifFuzz~\cite{DBLP:conf/icse/nilizadeh}.
			Legends: \#\textbf{R}: no. of methods,
			\#\textbf{S}$_F$: no. of secret values (\toolname),
			\#\textbf{P}$_F$: no. of public values (\toolname),
			$\varepsilon_{1}$: tolerance for $L_1$-norm functional clustering (\toolname),
			\#\textbf{K}$_{F}$: no. of clusters (\toolname),
			\textbf{Safe}$_F$: Yes, if there is only 1 cluster (\toolname),
			\#\textbf{S}$_D$: no. of secret values (DifFuzz),
			\#\textbf{P}$_D$: no. of public values (DifFuzz),
			$\delta$: max. cost difference (DifFuzz),
			\textbf{Safe}$_D$: Yes, if ${\delta \leq 1}$ (DifFuzz),
			\#\textbf{K}$_D$: no. of clusters (DifFuzz).
		}
		\label{tab:benchmark}
		\label{tab:benchmark-DifFuzz}
		 \resizebox{0.85\textwidth}{!}{
		\begin{tabular}{ | l  r | r  r  r  r  r | r  r  r  r  r |}
		\hline
		&   &  \multicolumn{5}{c|}{\toolname} & \multicolumn{5}{c|}
		{DifFuzz~\cite{DBLP:conf/icse/nilizadeh}} \\
		\cline{3-12}
		     	&      &    &    &    &    &   &   &   &   &      &   \\
			Benchmark & \#\textbf{R} & \#\textbf{S}$_F$ & \#\textbf{P}$_F$  &
			$\varepsilon_{1}$ & \textbf{Safe}$_F$ & \#\textbf{K}$_{F}$  &
			\#\textbf{S}$_D$ & \#\textbf{P}$_D$ &
			\textbf{Max. $\delta$} & \textbf{Safe}$_D$ & \#\textbf{K}$_D$ \\ \hline
			Zigzag  & 13 & 70 & 6,912 & 1 & No & 2 & 3,007 & 1,532 & 1 & Yes & 1  \\ \hline
			Guess\_Secret\_1 & 10 & 110 & 6,649 & 1 & No & 105 & 9,672 & 4,797 & 2 & No & 2 \\ \hline
			Guess\_Secret\_2 & 5 & 72 & 7,414 & 1 & No & 7 & 6,480 & 3,476 & 0 & Yes & 1 \\ \hline
			processBid  & 3 & 116 & 8,100 & 1 & No & 112 & 2,282  & 1,170 & 4  & No & 2 \\ \hline
			pwcheck\_unsafe  & 3 & 118 & 9,560 & 1 & No & 115	& 15,290 & 7,660 & 47 & No & 16 \\ \hline
			Branch\_and\_Loop\_1 & 4 & 179 & 1,761 & 1 & No & 4 & 8,303  & 4,477 & 30,404 & No & 4 \\ \hline
			Branch\_and\_Loop\_2 & 8 & 226 & 2,111 & 1 & No & 5 & 9,003  & 4,524 & 30,404 & No & 5 \\ \hline
			Branch\_and\_Loop\_3 & 16 & 224 & 2,101 & 1 & No & 7 & 4,419  & 2,556 & 30,405 & No & 6 \\ \hline
			Branch\_and\_Loop\_4 & 32 & 229 & 2,121 & 1 & No & 9 & 8,612 & 4,656 & 30,405 & No & 6 \\ \hline
			Branch\_and\_Loop\_5 & 64 & 238 & 2,213 & 1 & No & 19 & 10,523 & 5,337 & 30,405 & No & 7 \\ \hline
			Branch\_and\_Loop\_6 & 128 & 255 & 2,200 & 1 & No & 24 & 7,539 & 3,869 & 30,405 & No & 5 \\ \hline
		\end{tabular}
		}
		\vspace{-0.5em}
\end{table*}

\noindent\textbf{Clustering Comparison}.
Figure~\ref{fig:comp-kmeans-hierarichal} shows the comparison between
the hierarchical and constrained K-means algorithms
(Section~\ref{sec:implement}) using \texttt{Branch\_and\_Loop\_6} where
the number of secrets varies from 32 to 1,024 all with 1,000 public
values. It shows that the constrained K-means
is computationally expensive, while the hierarchical clustering is
much more scalable (up to $400\times$). Besides, the constrained K-means
discovers more clusters than the hierarchical one.
Note that the clusters discovered by both algorithms are
valid, and we prefer the one with the fewer number of clusters!
We use the hierarchical algorithm for the rest of this paper.

\noindent\textbf{Scalability}.
We examine the scalability of components in \toolname for fitting functions,
finding clusters, and learning decision trees.
We observe that \toolname can
handle more than 250 timing functions each defined
over more than 2,000 public values in less than 30 seconds.
The computation time grows in the quadratic factor with respect to the growth
of the number of secret values and public values.
Learning the decision tree model includes both fitting functions
over the auxiliary features and using CART algorithms in the
functional domain. We observe that this procedure is scalable
and  takes less than one minute in the worst case.

\noindent\textbf{DifFuzz Approach}.
DifFuzz is a recent side-channel detection
technique that outperforms other state-of-the-art
techniques~\cite{DBLP:conf/icse/nilizadeh}. The approach
extends AFL~\cite{AFL} and Kelinci~\cite{kersten2017poster}
fuzzers to detect side channels. The goal of DifFuzz is
to maximize the following objective: ${\delta = |c(p,s_1) - c(p,s_2)|}$,
that is, to find two distinct secret values $s_1,s_2$ and a public value $p$
that give the maximum cost ($c$) difference.
Because the cost function ($c$) is the number of executed bytecodes,
we use  BigInteger manipulations equivalent to
sleep commands in micro-benchmark programs.
Note that the objective function is based on
$\epsilon$-approximate noninterference where the goal
is to maximize the point-wise cost differences
between two secrets.

\noindent\textbf{DifFuzz versus \toolname}.
Table~\ref{tab:benchmark-DifFuzz} shows the results of applying
\toolname and DifFuzz on the set of micro-benchmark programs.
We generate inputs for both approaches in 30 mins.
We analyze the input generated from these two approaches
in two criteria: 1) whether they deem the benchmark safe?
2) how many classes of observations do they find?
Based on the results in~\cite{DBLP:conf/icse/nilizadeh},
the minimum value of cost difference $\delta$ for unsafe variants
is $8$. However, in this study, a program is safe if
${\delta \leq 1}$ that allows DifFuzz
to deem the application unsafe for smaller cost differences.
In the same way, we set the tolerance parameter based on $L_1$
norm distance in \toolname to be $1$.

We highlighted differences between DifFuzz
and \toolname in Table~\ref{tab:benchmark-DifFuzz}. First,
DifFuzz reports the max. cost difference in
\texttt{Zigzag} application as $1$, so the program is safe. This is largely
due to the point-wise noninterference definition in DifFuzz. The definitions with
the $L_1$-norm over the shapes can easily show higher costs
and deem the application unsafe.
Second, we apply the point-wise and functional
clusterings for inputs generated by DifFuzz and \toolname, respectively.
We observe that DifFuzz finds fewer clusters compared to \toolname.
There are mainly two reasons for these differences.
The first factor is due to the point-wise definitions
in finding classes of side channels as illustrated in
Section~\ref{func-side-channel-clusters}.
In \texttt{Guess\_1} program, for each distinct secret value,
there is a unique public value
where the execution time jumps from one cost
to another. These are captured by the functional clustering where
there is an almost equal number of secrets and clusters.
The second one is due to the objective function of DifFuzz that tries to find
two secret values (with the same public value) such that the cost differences between
them are maximized. \toolname, on the other hand, tries to find
as many functional clusters as possible.
This factor is the main reason for the differences
in \texttt{Branch\_and\_loop} applications.

\section{Case Studies}
\label{sec:case-study}
\begin{table*}[t]
	\caption{Case Studies. 
    Legends similar to
		Table~\ref{tab:benchmark}, except,
		\#\textbf{M} the number of methods
		in applications,
		$\varepsilon_{0,1}: $ tolerance for
		$L_1$-norm of the timing model,
		$\varepsilon_{1,1}: $ tolerance  for
		$L_1$-norm of the first derivative of the timing model,
		\textbf{A}: accuracy of the tree model,
		\textbf{H}: height of the tree,
		\#\textbf{L}: number of leaf nodes in the tree,
		\textbf{T}: computation time for decision tree learning (s).
	}
	\label{tab:summary}
	\resizebox{\textwidth}{!}{
		\begin{tabular}{ | l | r  r  r  r | r  r  r | r  r  r | r  r  r  r |}
			\hline
			&      &    &    &    &    &    &   &    &    &     &    &  &  &  \\
			Benchmark& \#\textbf{M} & \#\textbf{R} & \#\textbf{S} & \#\textbf{P} & $\varepsilon_{\scalebox{.6}{0,1}}$ & \#\textbf{K}$_{0,1}$ &
			\textbf{T}$_{0,1}$ & $\varepsilon_{\scalebox{.6}{1,1}}$ & \#\textbf{K}$_{1,1}$ & \textbf{T}$_{1,1}$ & \textbf{A} & \textbf{H} &
			\#\textbf{L} & \textbf{T}  \\ \hline
			Regex & 620 & 203 & 1,154 & 6,365 & 2e-1 & 162 & 1,801 & 2e-1 & 49 & 4,812 & 89.7\% & 14 & 120 & 2,084.0 \\ \hline
			Jetty & 63 & 56 & 800 & 635 & 1e-1  & 20  & 49.7 & 1e-2 & 15 & 82.6 & 99.4\% & 12 & 20 & 52.1 \\ \hline
			iControl (SOAP) & 41,541 & 127 & 342 & 1,164 & 1e-1  & 33 & 19.7 & 1e-1  & 19 &  43.2 &  98.2\% & 10 & 9 & 10.5  \\ \hline
			Javax (crypto) & 612 & 56 & 1,533  & 1,045 & 1e-1  & 54  & 174.2  &  1e-1  & 32  & 253.0 &  88.6\% & 6 & 36 & 7.2  \\ \hline
			GabFeed & 573 & 43 & 1,105 & 65 & 1e-1 & 34 & 58.5 & 1e-2 &  34 & 70.5 & 99.6\% & 31 & 34 & 41.7 \\ \hline
			Stegosaurus & 237 & 96 & 512 & 60 & 2e-1 & 5 & 3.6 & 1e-1 &  3 & 3.6 & 100.0\% & 4 & 5 & 12.6 \\ \hline
			SnapBuddy & 3,071 & 65 & 477 & 14 & 2e-1 & 13 & 2.8 & 2e-1 & 8 & 3.0 & 96.2\% & 14 & 13 & 3.1 \\ \hline
			ShareValue & 13 & 7 & 164 & 41 & 6e-2  & 29 & 0.7 & 1e-2 & 14 & 0.7 & 99.3\% & 17 & 29 & 3.4 \\ \hline
			MST(Kruskal) & 5 & 6 & 120 & 40 & 3e-1 & 20  & 0.4  & 3e-1 & 5 & 0.4  & 80\%  & 7  & 20  & 3.0  \\ \hline
			Collab & 185 & 53 & 176 & 11 & 1e-2  & 1 & 0.3 & 1e-2 & 1 & 0.3 & N/A & N/A & N/A & N/A  \\ \hline
		\end{tabular}
	}
\end{table*}

Table~\ref{tab:summary} summarizes $10$ Java applications used as
case studies. We consider
$L_1$-norm distance between timing functions
($\varepsilon_{\scalebox{.6}{0,1}}$) and their first derivatives
($\varepsilon_{\scalebox{.6}{1,1}}$).
The main research questions are
``Do functional clustering and decision tree learning (a) scale well and
(b) provide useful information about leaks?''

\noindent\textbf{A) Regex.}
Regex's case study was described in Section~\ref{sec:overview}.
To answer the research question:
\textit{Usefulness:} The decision tree pinpoints a location in
the regex package that leaks the value of secret patterns.
\textit{Scalability:} The overall computation time of clustering
and decision tree learning is about 65 mins.

\begin{figure*}[!t]
	\centering
	\includegraphics[width=0.28\textwidth]{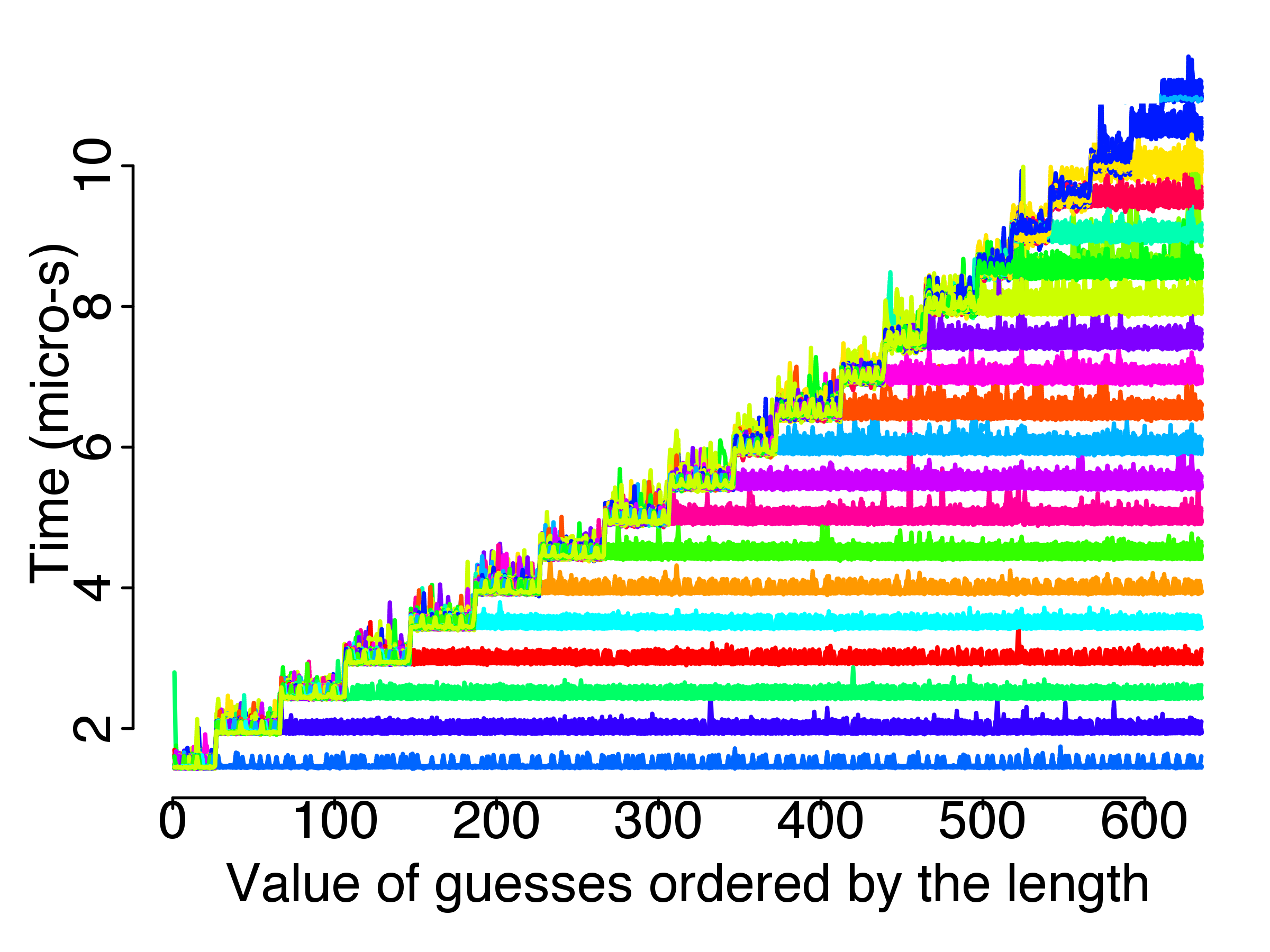}
  \hfill
  \scalebox{0.45}{
    \begin{tikzpicture}[align=center,node distance=1cm,->,thick,
        draw = black!60, fill=black!60]
      \centering
      \pgfsetarrowsend{latex}
      \pgfsetlinewidth{0.3ex}
      \pgfpathmoveto{\pgfpointorigin}

      \node[dtreenode,initial above,initial text={}] at (0,0) (l0)  {
        jetty.util.security.\\Credential.stringEquals\_bblock\_106};
      \node[dtreenode,below=of l0] (l2)
       {jetty.util.security.\\Credential.stringEquals\_bblock\_106};
       \node[dtreenode,below=of l2] (l4)
        {jetty.util.security.\\Credential.stringEquals\_bblock\_106};
        \node[below=of l4] (l6) {};

        \node[dtreeleaf,bicolor={blue!60 and blue!60 and 0.99},below left=of
          l0] (l1) {};
        \node[dtreeleaf,bicolor={blue and blue and 0.99},below left=of l2]
        (l3) {};
        \node[dtreeleaf,bicolor={green and green and 0.99},below left=of
          l4] (l5) {};

        \path[->]  (l0) edge  node [left,pos=0.4] {$ = \beta_1 ~~$} (l1);
        \path  (l0) edge  node [right, pos=0.4] {$~~ \neq \beta_1 $} (l2);
        \path  (l2) edge  node [left] {$ = \beta_2 ~~$} (l3);
        \path  (l2) edge  node [right] {$~~ \neq \beta_2 $} (l4);
        \path  (l4) edge  node [left] {$ = \beta_3 ~~$} (l5);
        \path  (l4) edge[dotted]  node [right] {$~~ \neq \beta_3$} (l6);
    \end{tikzpicture}
  }
	\hfill
	\includegraphics[width=0.28\textwidth]{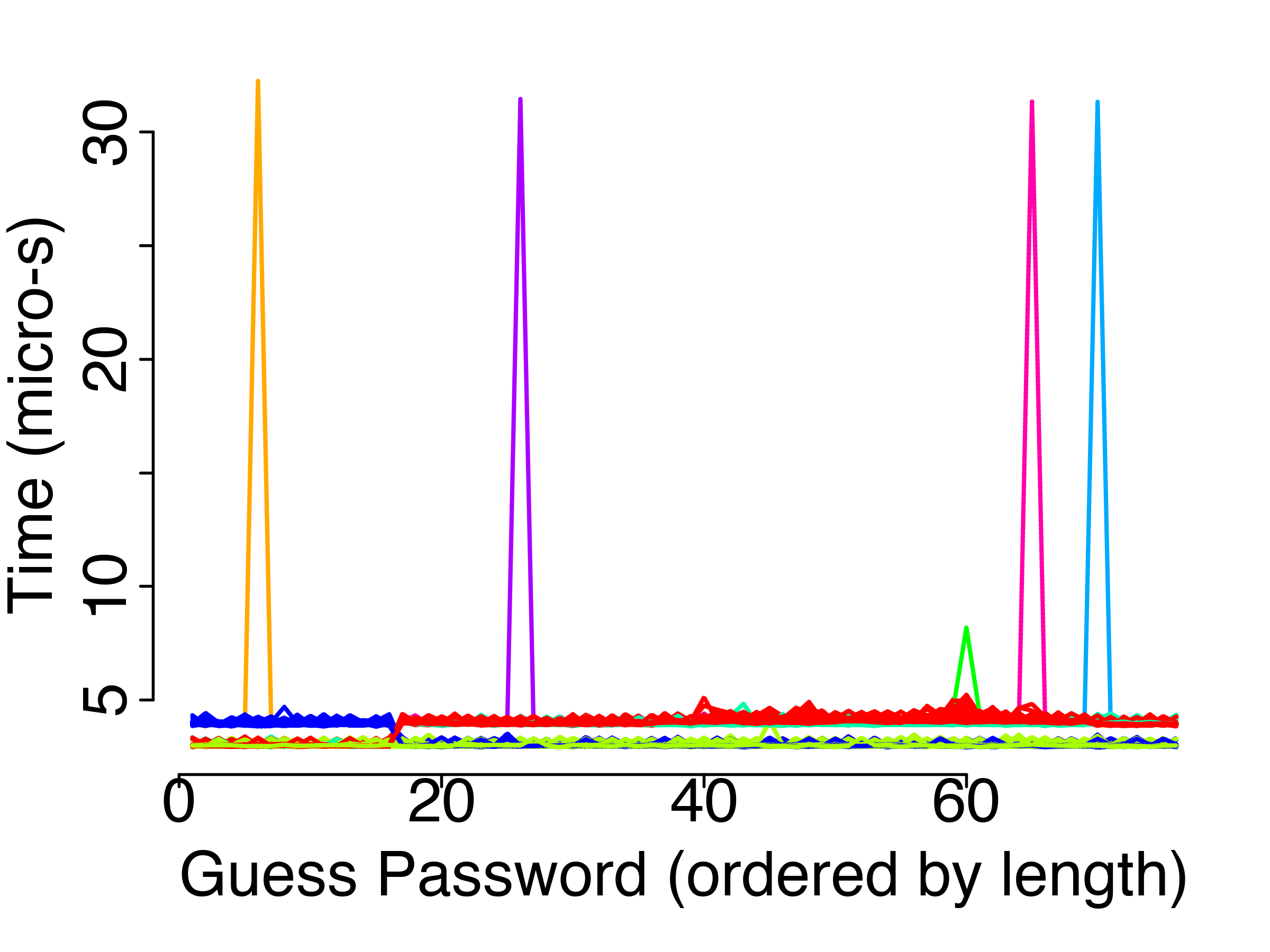}
	\hfill
	\scalebox{0.45}{
		\begin{tikzpicture}[align=center,node distance=1cm,->,thick,
				draw = black!60, fill=black!60]
			\centering
			\pgfsetarrowsend{latex}
			\pgfsetlinewidth{0.3ex}
			\pgfpathmoveto{\pgfpointorigin}

			\node[dtreenode,initial above,initial text={}] at (0,0) (l0)  {
				iControl.ManagementEventSubscription\\UserCredential.equals\_bblock\_95};
			\node[dtreenode,below=of l0] (l2)
				{iControl.ManagementEventSubscription\\UserCredential.equals\_bblock\_87};
			 \node[dtreenode,below=of l2] (l4)
				{iControl.ManagementEventSubscription\\UserCredential.equals\_bblock\_87};
				\node[below=of l4] (l6) {};

				\node[dtreeleaf,bicolor={brightgreen and brightgreen and 0.99},below left=of
					l0] (l1) {};
				\node[dtreeleaf,bicolor={blue and blue and 0.99},below left=of l2]
				(l3) {};
				\node[dtreeleaf,bicolor={mauve and mauve and 0.99},below left=of
					l4] (l5) {};

				\path[->]  (l0) edge  node [left,pos=0.4] {$ = 1 ~~$} (l1);
				\path  (l0) edge  node [right, pos=0.4] {$~~ = 0 $} (l2);
				\path  (l2) edge  node [left] {$ = \beta_1 ~~$} (l3);
				\path  (l2) edge  node [right] {$~~ \neq \beta_1 $} (l4);
				\path  (l4) edge  node [left] {$ = \beta_2 ~~$} (l5);
				\path  (l4) edge[dotted]  node [right] {$~~ \neq \beta_2$} (l6);
		\end{tikzpicture}
	}
	\caption{
		(a) $800$ Jetty timing functions are clustered into $20$ groups using
		$L_1$-norm. This indicates potential timing leaks in Jetty.
		What properties are leaking?
		(b) Jetty decision tree. The number of calls to the basic block at
		line 106 of stringEquals (shown in Figure~\ref{fig:jetty-streql})
    discriminates different clusters. The code region shows the length of
		secret passwords is leaking.
		(c) (parts of) $342$ iControl timing functions are clustered into $9$ groups.
		(d) (parts of) iControl decision tree model. It pinpoints basic
		blocks in \texttt{UserCredential.equals} method. The code
		region indicates that the whole secret password can be compromised with timing
		side channel attacks.
	}
	\label{fig:Jetty-clustered}
	\label{fig:Jetty-DT}
	\label{fig:iControl-clustered}
	\label{fig:iControl-DT}
\end{figure*}

\noindent\textbf{B) Jetty.}
We analyze the
\texttt{util.security} package of Eclipse Jetty web server. The package
has a \texttt{Credential} class which had a timing side channel. This
vulnerability was analyzed in \cite{DBLP:conf/ccs/ChenFD17} and fixed initially
in~\cite{jetty-1}. Then, the developers noticed that the implementation
in~\cite{jetty-1} can still leak information and fixed this issue with a
new implementation in~\cite{jetty-2}. We consider this new implementation
shown in Figure~\ref{fig:jetty-streql} and apply
\toolname to check its security. The final fix was done a few months
later~\cite{jetty-3}, but before we reported our finding to the developers.

\noindent\textit{Inputs}. The secret input is the password stored at the server,
and the public input is the guess. The defender starts by choosing a finite set of
secret and public values from the fuzzer.
The defender obtains 800 different secret passwords and 635 different guesses
from the fuzzer. The lengths of passwords are at most 20 characters.

\noindent\textit{Side Channel Discovery}.
For each secret value, \toolname varies 635 different guesses and
measures the execution time of Jetty. Then, \toolname
models the running time of 800 secret values with B-spline basis.
The next step is to find out how these functions are related based
on their functional distances.
Given the $L_{1}$-norm as the distance function and
the tolerance $\varepsilon = 0.1$,
\toolname uses the clustering algorithm
and returns $20$ classes of observations as shown in
Figure~\ref{fig:Jetty-clustered} (a).
The existence of $20$ distinct classes of observations indicates
the presence of a functional side channel in the Jetty package.

\begin{figure}[t!]
  \caption{String equality in
  Eclipse Jetty ({\tt s1} secret, {\tt s2} public).}
  \raggedright
  \begin{lrbox}{\mybox}%
    \begin{scriptsize}
      \begin{mylisting}[hbox,enhanced,drop shadow]{stringEquals}
boolean ?{\textbf{stringEquals}}?(String s1, String s2) {
  ?\indentrule?if (s1 == s2) return true;
  ?\indentrule?if (s1 == null || s2 == null) return false;
  ?\indentrule?boolean result = true;
  ?\indentrule?int l1 = s1.length(), l2 = s2.length();
  ?\indentrule?if (l1 != l2) result = false;
  ?\indentrule?int l = Math.min(l1, l2);
  ?\indentrule?for (int i = 0; i < l; ++i){    (line.106)
  ?\indentrule?  ?\indentrule?result &= (s1.charAt(i) == s2.charAt(i));}
  ?\indentrule?return result;
  ?\indentrule?}
      \end{mylisting}
    \end{scriptsize}
  \end{lrbox}%
  \scalebox{0.8}{\usebox{\mybox}}
  \label{fig:jetty-streql}
\end{figure}

\noindent\textit{Side Channel Explanation}.
Now, the defender wants to know what properties
of program internals leak through the timing side channels.
\toolname uses the instrumented Jetty
and obtains 56 internal features such as
method calls and basic block invocations. Each secret
value has the functional evaluation of 56 internal
features over the public inputs as well as a label from
the clustering. Next,
\toolname uses the decision tree inferences to localize
code regions that contribute to different observations.
Figure~\ref{fig:Jetty-DT} (b)
shows (parts of) the decision tree model learned for Jetty.
Using this model, the defender
realizes that different calls to a basic block in
{\tt Credential.StringEquals()} method are what
distinguishes the clusters.
This basic block represents the loop body of the {\tt for} statement
in the method shown in Figure~\ref{fig:jetty-streql}. For instance, the
green cluster (third from the bottom of the center diagram, the bottom of the right
diagram) corresponds to the case where {\tt stringEquals\_bblock\_106}
is executed according to $\beta_3$ function.
Note that edge values are B-spline
functions over public values, and the max value for
$\beta_i$ function is $i$. For the example of $\beta_3$, the
max value is $3$, and if the basic block for a secret value is
called at most three times, it belongs to the green cluster.
Using the decision tree model and the relevant code,
the defender realizes that the minimum of the
lengths (the secret and the guess) is leaking through the
calls to {\tt stringEquals\_bblock\_106}.

\noindent \textit{Usefulness:} The decision tree pinpoints a location in
Jetty that leaks the length of secrets.
\noindent \textit{Scalability:} The overall computation time is about 2 mins.

\noindent\textbf{C) iControl (SOAP).}
iControl (SOAP)\footnote{\url{https://clouddocs.f5.com/api/icontrol-soap/}}
is an open source API that uses SOAP/XML to establish communications between
dissimilar systems. The library has 41,541 methods.
One key confidentiality-related functionality of the library is to store credentials of various users and to validate their credentials against a given guess.
The defender's goal is to find out whether there exist timing side channels in the
library, and if so, identify the code regions potentially responsible for creating the side channels.

\noindent\textit{Inputs}. The natural candidate for the secret input in this application is the stored credential at the server, while the public input is a given guess against a stored credential. The defender considers the lexicographic ordering over public inputs to generate functional data. With a timeout of two hours on fuzzing,
the defender obtains $342$ unique secret credentials and $1,164$ unique guesses.
The credentials include secret passwords with the length of at most 16 characters.

\noindent\textit{Side Channel Discovery}.
\toolname begins by varying $1,164$ guesses for each secret
input and uses B-spline to model $342$ timing functions. With the default parameters
($L_1$-norm as the distance norm and $\varepsilon = 0.1$) for the clustering,  \toolname identifies $33$ classes of observations. Figure~\ref{fig:iControl-clustered} (c) shows timing functions and corresponding clusters.
The presence of multiple clusters points towards the existence of timing side channels.

\noindent\textit{Side Channel Explanation}.
The defender specifies basic block calls as the features to be used in the explanation of the side channels.
\toolname runs previously identified inputs on the instrumented version of iControl and generates $127$ auxiliary features (basic blocks) about the internals of iControl.
Given the set of traces containing the information on these features and corresponding cluster labels, \toolname uses decision tree models to present an explanation for the side channels.
The decision tree is shown in
Figure~\ref{fig:iControl-DT} (d). It pinpoints that
different calls to the basic block at the
equality check in \texttt{ManagementEventSubscriptionUserCredential} class
is a potential explanation of the timing differences. Using this information and
the relevant code, the defender may infer that the application uses Java
string equality check to compare passwords. This leads to a
password-matching style vulnerability where an attacker can obtain a prefix
of secret passwords in each step of attack.

We reported this vulnerability to both F5 security team and F5 open-source community developers.
The F5 security team has confirmed this vulnerability. Moreover, this explanation helped them to
identify a potential vulnerability in their closed-source implementations.

\begin{figure*}[!t]
	\centering
	\includegraphics[width=0.28\textwidth]{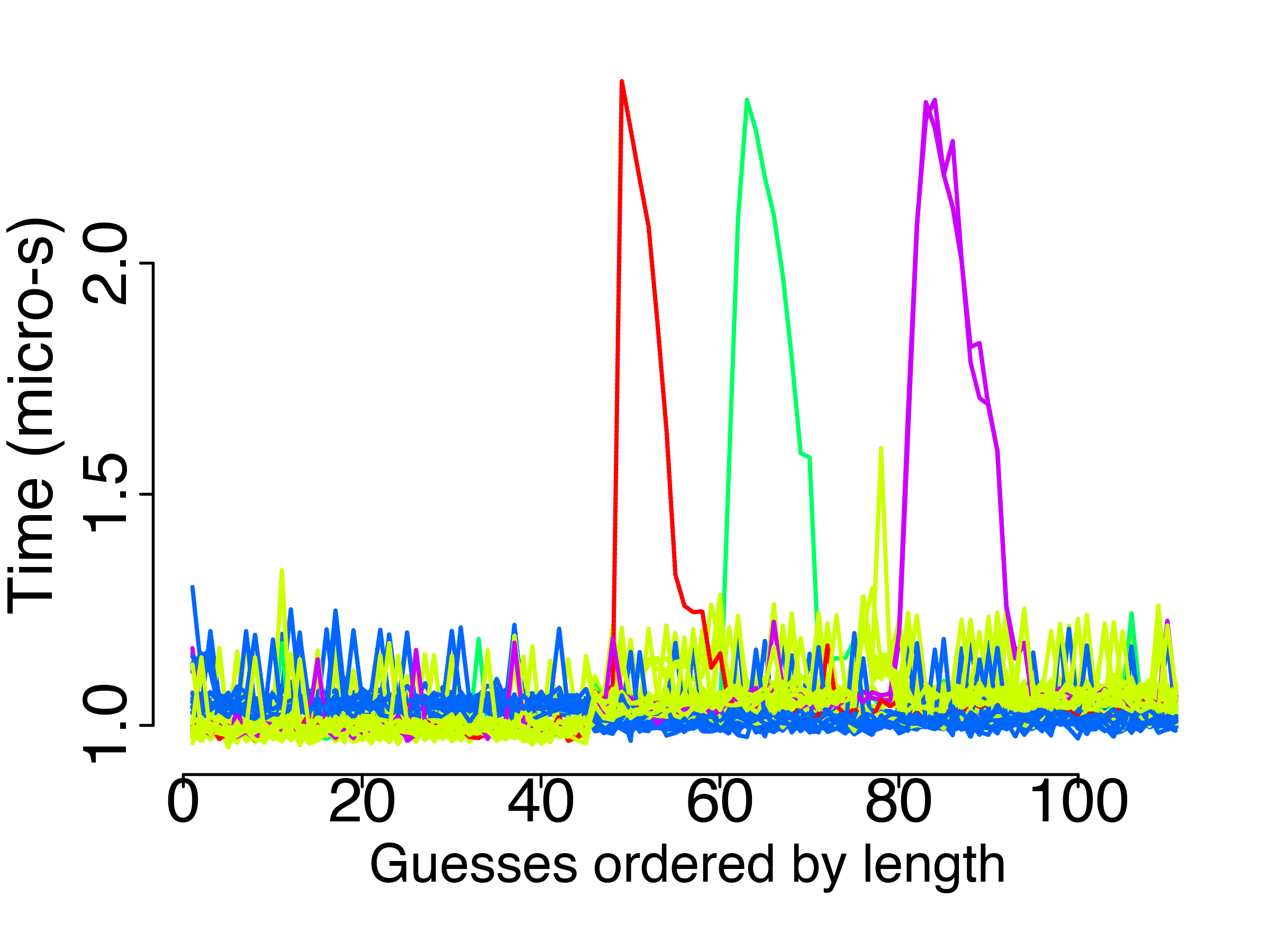}
	\hfill
	\scalebox{0.45}{
		\begin{tikzpicture}[align=center,node distance=1cm,->,thick,
				draw = black!60, fill=black!60]
			\centering
			\pgfsetarrowsend{latex}
			\pgfsetlinewidth{0.3ex}
			\pgfpathmoveto{\pgfpointorigin}

			\node[dtreenode,initial above,initial text={}] at (0,0) (l0)  {
				javax.crypto.spec\\.SecretKeySpec.equals\_bblock\_245};
			\node[dtreenode,below=of l0] (l2)
			 {javax.crypto.spec\\.SecretKeySpec.equals\_bblock\_245};
			 \node[dtreenode,below=of l2] (l4)
				{javax.crypto.spec\\.SecretKeySpec.equals\_bblock\_245};
				\node[below=of l4] (l6) {};

				\node[dtreeleaf,bicolor={green and green and 0.99},below left=of
					l0] (l1) {};
				\node[dtreeleaf,bicolor={red and red and 0.99},below left=of l2]
				(l3) {};
				\node[dtreeleaf,bicolor={mauve and mauve and 0.99},below left=of
					l4] (l5) {};

				\path[->]  (l0) edge  node [left,pos=0.4] {$ = \beta_1 ~~$} (l1);
				\path  (l0) edge  node [right, pos=0.4] {$~~ \neq \beta_1 $} (l2);
				\path  (l2) edge  node [left] {$ = \beta_2 ~~$} (l3);
				\path  (l2) edge  node [right] {$~~ \neq \beta_2 $} (l4);
				\path  (l4) edge  node [left] {$ = \beta_3 ~~$} (l5);
				\path  (l4) edge[dotted]  node [right] {$~~ \neq \beta_3$} (l6);
		\end{tikzpicture}
	}
	\hfill
	\includegraphics[width=0.28\textwidth]{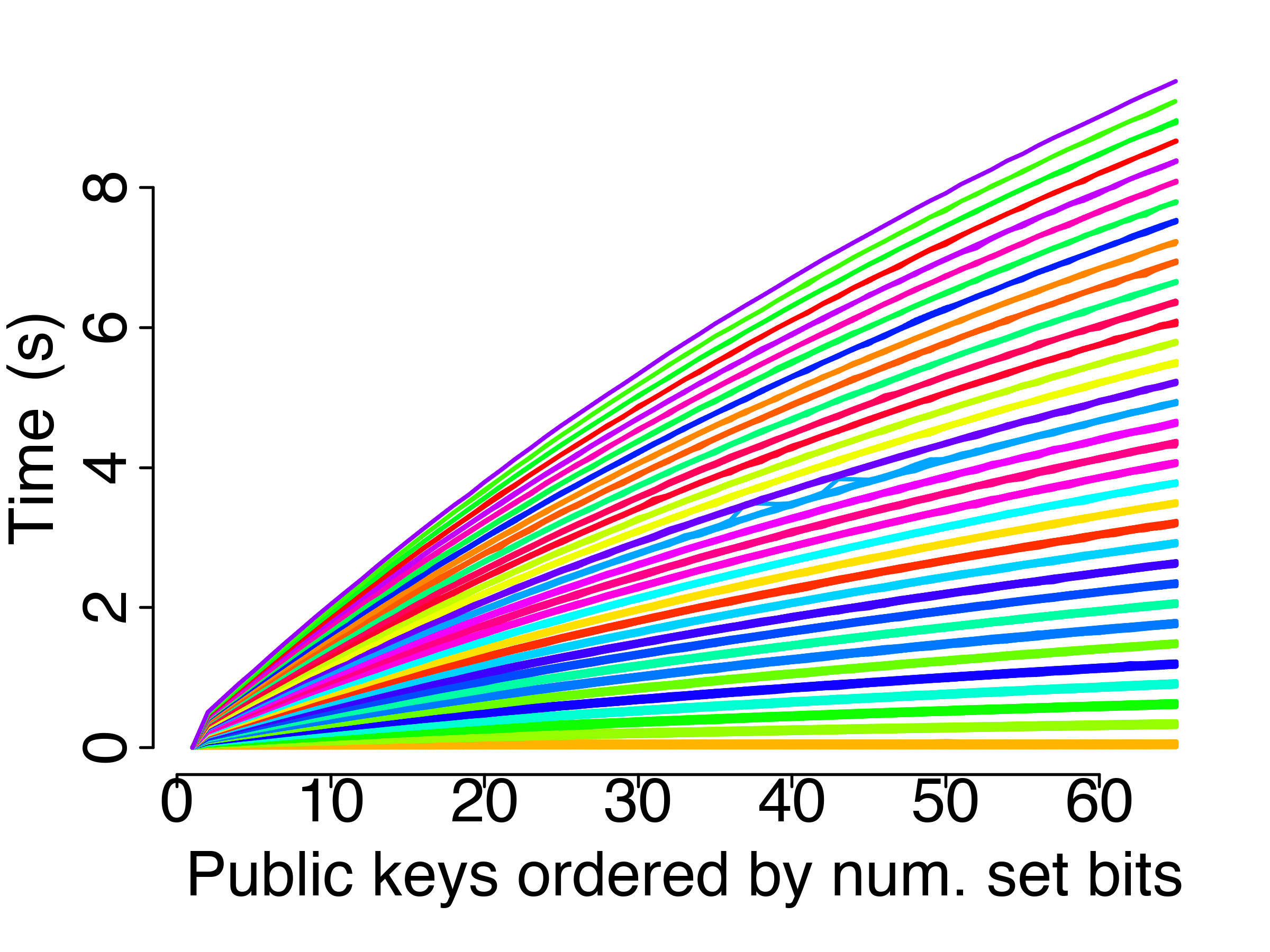}
	\scalebox{0.45}{
	\begin{tikzpicture}[align=center,node distance=0.8cm,->,thick,
		draw = black!60, fill=black!60]
		\centering
		\pgfsetarrowsend{latex}
		\pgfsetlinewidth{0.3ex}
		\pgfpathmoveto{\pgfpointorigin}

		\node[dtreenode,initial above,initial text={}] at (5,0) (l0)
		{OptimizedMultiplier.standard\\Multiply\_BasicBlock\_18};
		\node[dtreenode,below=of l0] (l2)
		{OptimizedMultiplier.standard\\Multiply\_BasicBlock\_18};
		\node[dtreenode,below=of l2] (l4)
		{OptimizedMultiplier.standard\\Multiply\_BasicBlock\_18};
		\node[below=of l4] (l6) {};

		\node[dtreeleaf,bicolor={amber and amber and 0.99},below left=of l0] (l1) {};
		\node[dtreeleaf,bicolor={brightgreen and brightgreen and 0.99},below left=of l2]  (l3) {};
		\node[dtreeleaf,bicolor={brightgreen_2 and brightgreen_2 and 0.99},below left=of l4] (l5) {};

		\path[->]  (l0) edge  node [left,pos=0.4] {$ = 3*y ~~$} (l1);
		\path  (l0) edge  node [right, pos=0.4] {$~~ \neq 3*y  $} (l2);
		\path  (l2) edge  node [left] {$ = 127*y ~~$} (l3);
		\path  (l2) edge  node [right] {$~~ \neq 127*y $} (l4);
		\path  (l4) edge  node [left] {$ = 251*y ~~$} (l5);
		\path  (l4) edge[dotted]  node [right] {$~~ \neq 251*y$} (l6);
	\end{tikzpicture}
	}

	\caption{
		(a) (parts of) $1,533$ timing functions of javax Crypto are clustered
		into $5$ groups. (b) (parts of) javax Crypto decision tree model. It localizes
		the basic blocks in \texttt{SecretKeySpec.equals} that can leak the
		entire secret key due to the use of Java internals to compare byte arrays.
		(c) $1,105$ GabFeed timing functions are clustered into $34$ groups.
		(d) GabFeed decision tree shows the basic block at line 18
		of \texttt{standardMultiply} method is the discriminants.
		The code region shows the number of set bits in the secret key is leaking.
	}
	\label{fig:JavaX-clustered}
	\label{fig:Javax-DT}
	\label{fig:GF-clustered}
	\label{fig:GF-DT}

\end{figure*}

\noindent\textbf{D) Javax Crypto.}
Javax library provides the classes and interfaces for cryptographic operations in Java.
The \texttt{crypto} package in the library has 612 methods and provides
functionalities such as creating and modifying symmetric secret keys.
We analyzed the \texttt{crypto} package of javax library\footnote{\url{https://hg.openjdk.java.net/jdk8/jdk8/jdk/file/687fd7c7986d/src/share/classes/javax/crypto}} against timing side-channel vulnerabilities.

\noindent\textit{Inputs}.
The secret input is the symmetric secret key of encryption
algorithms (such as ``DES''), and the public input is a guess key to be compared against
the secret key. During two hours of fuzzing, the defender generates 1,533 secret keys
and 1,045 guess keys. The length of a key is at most 16 bytes.

\noindent\textit{Side Channel Discovery}.
\toolname identified 1,533 timing functions
 using B-spline basis and
returned $54$ clusters, as shown in
Figure~\ref{fig:JavaX-clustered} (a), with default parameters ($L_1$-norm and $\varepsilon = 0.1$).
The presence of $54$ classes of observations indicates the existence of timing
side channels.

\noindent\textit{Side Channel Explanation}.
The next step is to identify the culprit code regions
and understand what properties of secret keys are leaking.
\toolname runs the same set of
secret and guess inputs over the instrumented version (to output information
about the basic blocks) of the crypto library.
This results in generating $56$ auxiliary features about the
basic block calls. Given the basic block evaluations
and the cluster for each secret value, the decision tree model explains
which basic blocks contribute to different timing observations.
The decision tree in Figure~\ref{fig:Javax-DT} (b) shows the calls to a basic block
in \texttt{spec.SecretKeySpec.equals()} method is the root cause of
timing side channels:

\begin{lrbox}{\mybox}%
	 \raggedright
	\begin{scriptsize}
		\begin{mylisting}[hbox,enhanced,drop shadow]{SecretKeySpec.equals(Object obj)}
?\indentrule?if (this == obj)
?\indentrule?   return true;
?\indentrule?if (!(obj instanceof SecretKey))
?\indentrule?   return false;
?\indentrule?String thatAlg = ((SecretKey)obj).getAlgorithm();
?\indentrule?if (!(thatAlg.equalsIgnoreCase(this.algorithm))) {
?\indentrule?   ...
?\indentrule?}
?\indentrule?byte[] thatKey = ((SecretKey)obj).getEncoded();
?\indentrule?return java.util.Arrays.equals(this.key, thatKey);
		\end{mylisting}
	\end{scriptsize}
\end{lrbox}%
\scalebox{0.8}{\usebox{\mybox}}
This results in calling to \texttt{util.Arrays.equals()}:

\begin{lrbox}{\mybox}%
	 \raggedright
	\begin{scriptsize}
		\begin{mylisting}[hbox,enhanced,drop shadow]{Arrays.equals(byte[] a, byte[] a2)}
?\indentrule?if (a==a2)  return true;
?\indentrule?if (a==null || a2==null) return false;
?\indentrule?int length = a.length;
?\indentrule?if (a2.length != length) return false;
?\indentrule?for (int i=0; i<length; i++)
?\indentrule?   if (a[i] != a2[i]) return false;
?\indentrule?return true;
		\end{mylisting}
	\end{scriptsize}
\end{lrbox}%
\scalebox{0.8}{\usebox{\mybox}}

This internal method for the equality check of byte arrays is vulnerable
to timing side-channel attacks. The method returns as soon as there is a mismatch
between two byte arrays. An attacker can exploit this
vulnerability to recover secret keys.

We reported this problem to OpenJDK security
team. During the discussion, we were informed that the vulnerability has since been
fixed in an updated version of JDK-8~\cite{fix1}
(we analyzed JDK-8 project, while the fix appears in JDK-8-u project).
We also analyzed the implementations in JDK-8-u project~\cite{fix1} with the
same set of inputs from the previous analysis. During this analysis, we found out
that there are $7$ clusters in timing observations. This shows that the new implementation
has not completely fixed the side channels. The decision tree explains
that there are different calls to the basic block
at line $454$ in \texttt{isEqual()} method of \texttt{MessageDigest} class
~\cite{fix2}. Looking into the source code, we observed that the length of secret
byte arrays is leaking via timing side channels. Furthermore, the vulnerability
applies to any functionalities in javax that compare byte arrays.
We reported this vulnerability to the developers and suggested safe implementations
to fix it.

\begin{figure*}[!t]
	\centering
	\includegraphics[width=0.27\textwidth]{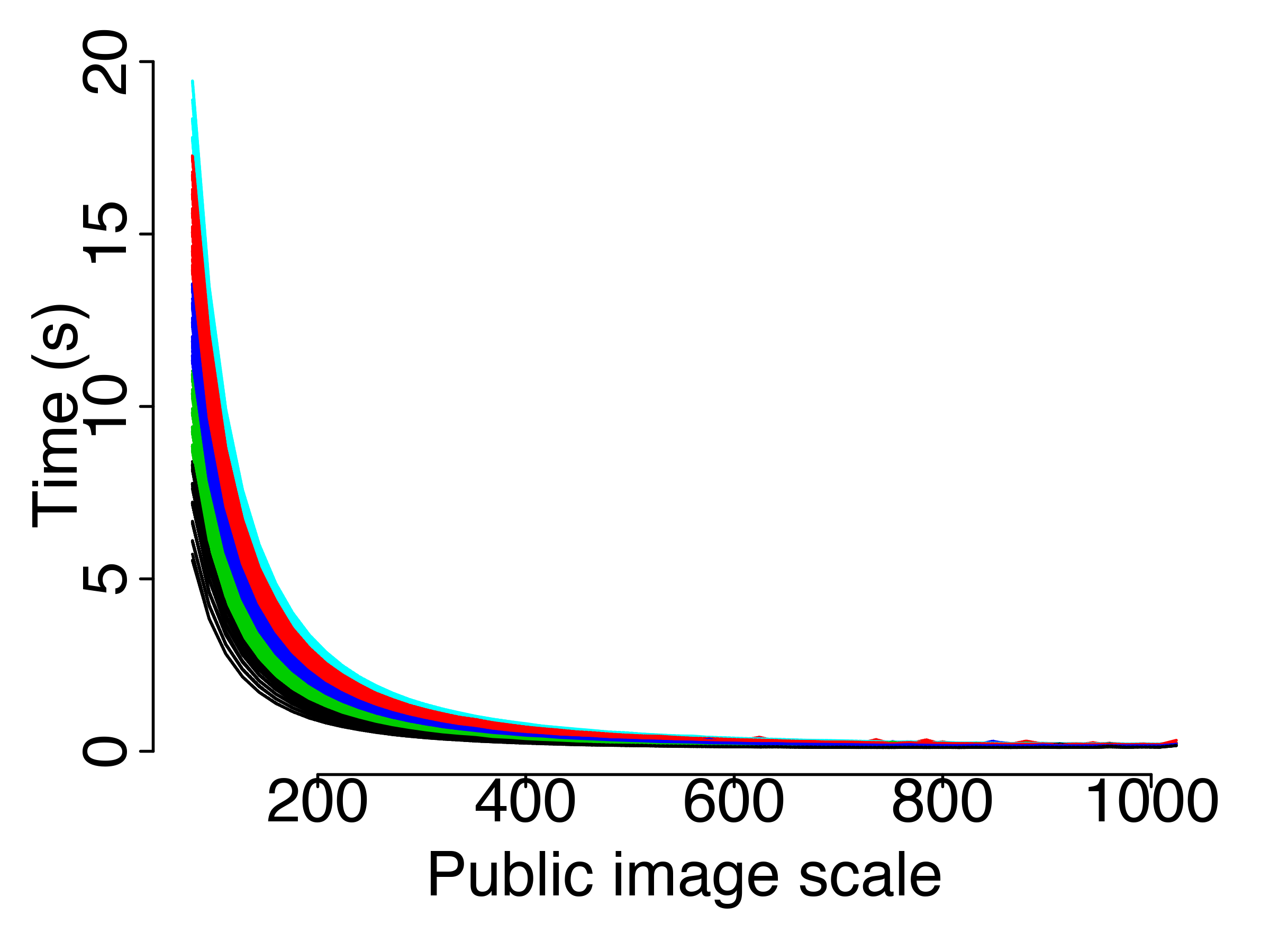}
	\scalebox{0.45}{
	\begin{tikzpicture}[align=center,node distance=0.8cm,->,thick,
		draw = black!60, fill=black!60]
		\centering
		\pgfsetarrowsend{latex}
		\pgfsetlinewidth{0.3ex}
		\pgfpathmoveto{\pgfpointorigin}

		\node[dtreenode,initial above,initial text={}] at (5,0) (l0)
		{com.bbnStegger.\\hide\_BasicBlock\_145};
		\node[dtreenode,below=of l0] (l2)
		{com.bbnStegger.\\hide\_BasicBlock\_145};
		\node[dtreenode,below=of l2] (l4)
		{com.bbnStegger.\\hide\_BasicBlock\_145};
		\node[below=of l4] (l6) {};

		\node[dtreeleaf,bicolor={black and black and 0.99},below left=of l0] (l1) {};
		\node[dtreeleaf,bicolor={green and green and 0.99},below left=of l2]  (l3) {};
		\node[dtreeleaf,bicolor={blue and blue and 0.99},below left=of l4] (l5) {};

		\path[->]  (l0) edge  node [left,pos=0.4] {$ <= 5.5E7*y^{-2} ~~$} (l1);
		\path  (l0) edge  node [right, pos=0.4] {$~~ > 5.5E7*y^{-2}  $} (l2);
		\path  (l2) edge  node [left] {$ <= 7.0E7*y^{-2} ~~$} (l3);
		\path  (l2) edge  node [right] {$~~ > 7.0E7*y^{-2} $} (l4);
		\path  (l4) edge  node [left] {$ <= 8.5E7*y^{-2} ~~$} (l5);
		\path  (l4) edge[dotted]  node [right] {$~~ > 8.5E7*y^{-2}  $} (l6);
	\end{tikzpicture}
	}
	\hfill
	\includegraphics[width=0.27\textwidth]{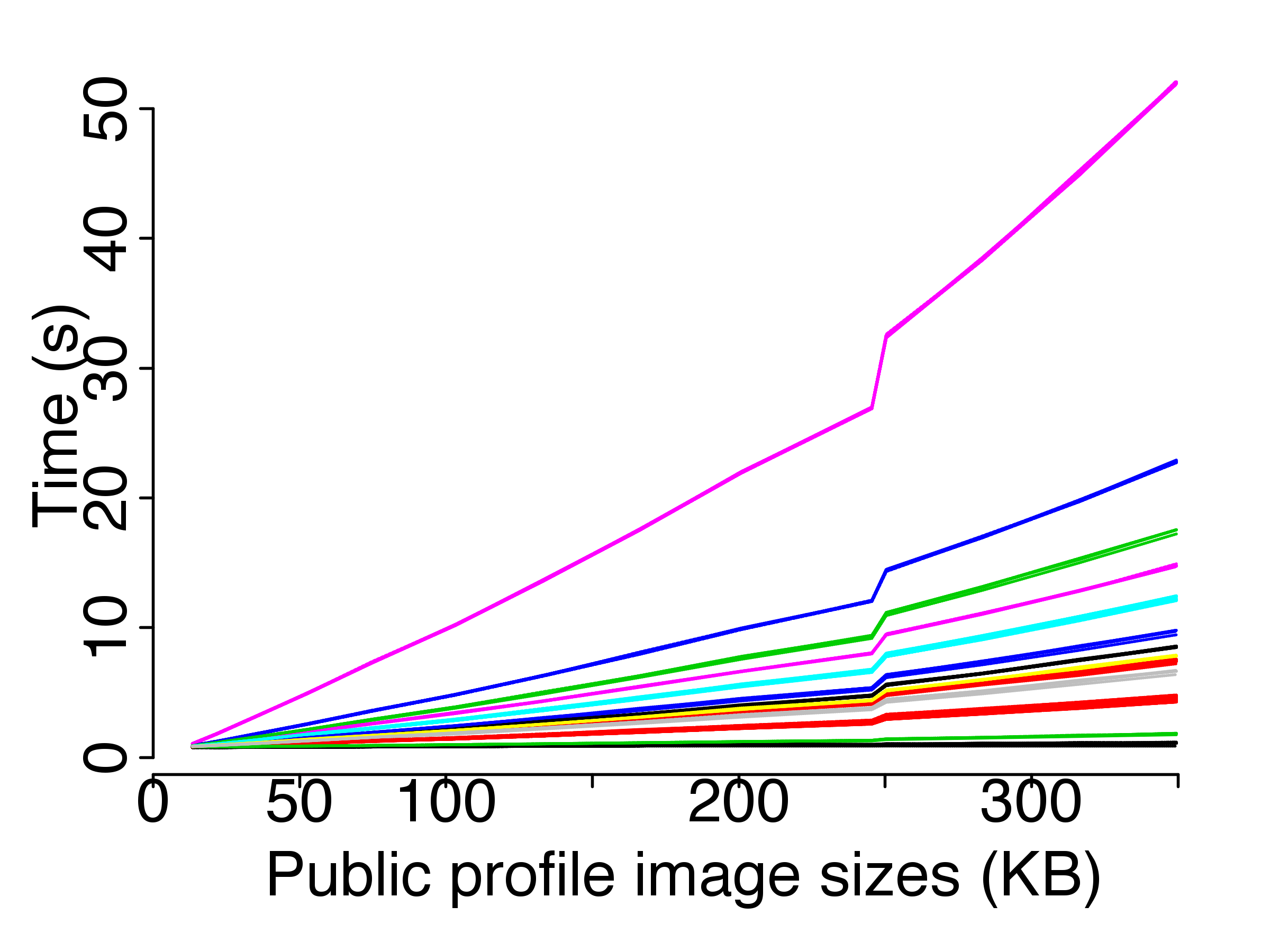}
	\scalebox{0.3}{
		\begin{tikzpicture}[align=center,node distance=0.8cm,->,thick,
		draw = black!60, fill=black!60]
		\centering
		\pgfsetarrowsend{latex}
		\pgfsetlinewidth{0.3ex}
		\pgfpathmoveto{\pgfpointorigin}

		\node[dtreenode,initial above,initial text={}] at (0,0) (l0)  {
			model.Filter.\\filter};
		\node[dtreenode,below right=of l0] (l2)
		{image.OilFilter.\\filterPixels};
		\node[dtreenode,below left=of l2] (l4)
		{image.ChromeFilter.\\access};
		\node[dtreenode,below left=of l4] (l6)
		{image.AbstractBuffered\\ImageOp.getRGB};
		\node[dtreenode,below left=of l6] (l7)
		{image.BicubicScaling\\Filter.filter};
		\node[dtreenode,below right=of l6] (l8)
		{image.FlipFilter\\.filter};

		\node[dtreeleaf,bicolor={black and black and 0.99},below left=of l0] (l1) {};
		\node[dtreeleaf,bicolor={pink and pink and 0.99},below right=of l2]  (l3) {};
		\node[dtreeleaf,bicolor={blue and blue and 0.99},below right=of l4]  (l5) {};
		\node[below left=of l7]  (empty3) {};
		\node[below right=of l8]  (empty4) {};
		\node[below right=of l7]    (empty1){};
		\node[below left=of l8]     (empty2){};

		\path[->]  (l0) edge  node [left,pos=0.4] {$ = 0 ~~$} (l1);
		\path  (l0) edge  node [right, pos=0.4] {$~~ = 1 $} (l2);
		\path  (l2) edge  node [left] {$ = 1 $} (l3);
		\path  (l2) edge  node [right] {$~~ = 0 $} (l4);
		\path  (l4) edge  node [right] {$= 1 ~~$} (l5);
		\path  (l4) edge  node [left] {$~~ = 0$} (l6);
		\path  (l6) edge  node [left] {$= 0 ~~$} (l7);
		\path  (l6) edge  node [right] {$~~ = 1$} (l8);
		\path  (l7) edge[dotted]  node [left] {$= 1 ~~$} (empty3);
		\path  (l7) edge[dotted]  node [right] {$~~ = 0$} (empty1);
		\path  (l8) edge[dotted]  node [left] {$~~ = 0$} (empty2);
		\path  (l8) edge[dotted] node [right] {$~~ = 1$} (empty4);
		\end{tikzpicture}
	}
	\caption{
		(a) $512$ Stegosaurus timing functions are clustered into $5$ groups.
		(b) Stegosaurus decision tree model. It pinpoints the basic
		block at line 145 of the \texttt{hide} method. The code
		region indicates that the length of secret messages is leaking.
		(c) $477$ timing functions of users' profiles in SnapBuddy are clustered
		into $13$ groups. (d) SnapBuddy decision tree model: calls to photo filter functions
		are discriminants. The type of photo filters applied
		by users on their public profiles can leak their identities.
		}
		\label{fig:ST-clustered}
		\label{fig:ST-DT}
	\label{fig:SB-clustered}
	\label{fig:SB-DT}
\end{figure*}

\noindent\textbf{E) GabFeed.}
GabFeed is a Java application with 573 methods implementing a chat server
~\cite{DBLP:conf/ccs/ChenFD17}.

\noindent\textit{Inputs}.
The server takes users' public key and its own private key
to generate a common key.
The defender uses \toolname to obtain 1,105 server's private keys
and 65 public keys where the public keys are ordered by their number
of set bits. In total, there are 71,825 test cases.

\noindent\textit{Side Channel Discovery}. For each secret key, \toolname varies public
keys and measures the execution time to generate the common key.
Next, \toolname uses B-spline and creates timing functions
for each secret. The next step is to find the equivalence relations
over the secret input using the functional clustering.
The defender provides $L_1$-norm and $\varepsilon_{\scalebox{.6}{0,1}}=0.1$ as parameters to
the hierarchal clustering, and \toolname discovers $34$ clusters
as shown in Figure~\ref{fig:GF-clustered} (c). This step partitions
1,105 secret values into $34$ distinguishable classes.

\noindent\textit{Side Channel Explanation}.
The next step is to find out what properties of secret keys are leaking.
\toolname runs the same secret and public inputs over
the instrumented GabFeed and obtains $43$ different auxiliary features.
Given the basic block evaluations
and the cluster for each secret value,
the task is to learn what basic blocks contribute to different clusters.
\toolname uses the CART algorithm and
produces the model in Figure~\ref{fig:GF-DT} (d).
Using this model, the defender observes that the number
of basic block calls at line 18
of the \texttt{standardMultiply} method explains
different clusters.
The edge values in the decision tree model is a linear function of
the public input with different slopes.
Inspecting the source code, the basic block executes expensive
shift left and add operations over BigIntegers of public keys for
each set bit in the secret key. The slopes in the edge values of the
decision tree model depend on the number of set bits in the secret key.

\noindent\textit{Usefulness:} The decision tree explains that
the calls to an expensive basic block is a linear
function of public key where the slope depends on the number of
set bits in secret key. GabFeed authentication algorithm leaks
set bits in the secret.
\noindent\textit{Scalability:} The overall computation time is about 2 mins.

\noindent\textbf{F) Stegosaurus.}
Stegosaurus with 237 methods is a messaging service
that uses steganographic algorithms to hide secret messages~\cite{Stegosaurus}.
The application takes the secret message with a $128$-bit key and
embeds the message in a random image.

\noindent\textit{Inputs}.
The secrets are the message
and the key. The defender uses \toolname
to generate $512$ secret messages with a size of at most $8$ letters.
We assume that the secret key is a fixed random value
chosen by the service. The public input is an image chosen by
the defender and ordered by their scales.

\noindent\textit{Side Channel Discovery}.
For each secret message, \toolname varies the scale of
images from 80$\times$80 to 1,024$\times$1,024. Then, it
measures the execution time of the application to
encode the message in different images. In total,
\toolname models $512$ timing functions. The next step is to
find the equivalence classes over these functions.
The defender provides $L_1$-norm and
$\varepsilon_{\scalebox{.6}{0,1}}=0.2$
as parameters to the clustering, and \toolname finds
$5$ classes of observations shown in
Figure~\ref{fig:ST-clustered} (a).

\noindent\textit{Side Channel Explanation}.
The next step is to find out what properties of secret
messages are leaking.
\toolname uses the instrumented version of Stegosaurus
to generate basic block calls for each execution.
In total, \toolname obtains 96 different basic block
calls each as a function over public inputs.
Given the basic block calls and the (timing)
cluster of each secret value,
\toolname uses the CART inference
to explain which of 96 basic blocks
contributes to different observations.
The decision tree model is shown in Figure~\ref{fig:ST-DT} (b).
The defender realizes that
the number of basic block calls at line 145
of the \texttt{hide} method explains
different clusters:

\begin{lrbox}{\mybox}%
	 \raggedright
	\begin{scriptsize}
		\begin{mylisting}[hbox,enhanced,drop shadow]{hide()}
?\indentrule?for (int pos=0; pos<=message.length(); ++pos) {
?\indentrule?   ...
?\indentrule?   while (pkCopy.compareTo(maxOffset) > 0) {
?\indentrule?(l.145)   pkCopy = pkCopy.subtract(maxOffset.multiply(perf));
?\indentrule?   }
?\indentrule?   ...}
		\end{mylisting}
	\end{scriptsize}
\end{lrbox}%
\scalebox{0.72}{\usebox{\mybox}}

\noindent In the above code snippet, {\tt pkCopy} is the fixed secret key, {\tt perf}
is a constant BigInteger value, and {\tt maxOffset} is the BigInteger representation
of the image scale (height$\times$width).

\noindent\textit{Usefulness:} The decision tree model shows the number of
calls to the basic block at line $145$ depends
on the secret message length directly and the scale of the image inversely.
Thus, the length of secret messages is the leaking property.
\noindent\textit{Scalability:} The overall computation time is less than 1 min.

\noindent\textbf{G) SnapBuddy.}
SnapBuddy with 3,071 methods is a mock social network where each user
has their own page with a photograph~\cite{snapbuddy}.
The size of profiles is a public input (observable through
the generated packets), and the identity of users actively interacting
with the server is a secret input.

\noindent\textit{Inputs}.
The defender considers the identities of 477
users currently in the network as
the secret inputs and varies the size of public profiles
from 13KB to 350KB.

\noindent\textit{Side Channel Discovery}.
\toolname uses B-spline to model the profile retrieval times for each
user as a function of profile sizes. \toolname models 477
timing functions, one for each user.
The next step is to find out the relationships
between the timing functions of different users and determine
if there are timing side channels. For this aim, \toolname applies the
clustering algorithm to identify different classes of observations.
\toolname discovers 13 clusters ($\varepsilon_{\scalebox{.6}{0,1}}$=0.2)
shown in Figure~\ref{fig:SB-clustered} (c). The clustering
partitions timing observations for 477 users into 13 equivalence
classes.

\noindent\textit{Side Channel Explanation}.
The next step is to find out what properties of users' public
profiles are leaking. In this example, in particular, it is difficult
to find out the leaking property solely based on the profile features
since it is exhaustively large. Some examples are the users' locations,
their names, their friends, their friends' name, their friends' location,
to mention but few.
This is one reason that we turn into collecting
program internal features through instrumentations.
The instrumentation provides 65 auxiliary features, and we model
them as functions over the profile sizes.
Figure~\ref{fig:SB-DT} (d) shows (part of) the decision
tree model that says users who do not apply
any filter on their images follow the black cluster (the bottom cluster in
Figure~\ref{fig:SB-clustered} (c)), while those who apply
oilFilter on their images are assigned to the pink cluster (the top cluster in
Figure~\ref{fig:SB-clustered} (c)). The decision tree shows that
it is the type of photo filters applied by the users on their public profile images
that are leaking. A passive attacker can use this
information to reduce her uncertainty about the identity of a user whom
the server downloads his/her profile, especially if some filters used by
a few users in the SnapBuddy.

\noindent\textit{Usefulness:}
The decision tree model explains non-trivial facts about leaks.
It shows that different photo filters applied by users on their profiles are leaking.
The defender can use this information to debug timing differences related to
the image filters.
\textit{Scalability:} The overall analysis takes less than 1 min.

\noindent\textbf{H) Share Value.}
The application is an extension of
classical share value program studied in
\cite{agat2000transforming,mantel2015transforming}.
In this case, every user in the system has
public and private shares. The program calculates
useful statistics about shares.

\noindent\textit{Inputs}. The program has 164 users
each with maximum of 400 private shares. The user can
have 1 to 400 public shares.

\noindent\textit{Side Channel Discovery}.
\toolname generates private and public shares in the given range.
In particular, it fixes private shares for each user.
Next, \toolname varies the number of public shares for each user
and measure the response times to calculate the statistics for each user.
Then, it fits B-spline to the execution times
and applies the functional clustering that
discovers 29 clusters with $\varepsilon_{\scalebox{.6}{0,1}}=0.06$.

\noindent\textit{Side Channel Explanation}.
The next step is to find out what properties of private shares
are leaking by using richer information from program internals.
The decision tree model shows that different intervals of calls to connect
to a remote database is the root cause of the leaks.
Therefore, the number of secret shares are leaking
through the time required to connect a remote DB.
The overall analysis takes about 4 (s).

\noindent\textbf{I) Kruskal.}
We analyze Kruskal's algorithm~\cite{kruskal1956shortest}
and its implementation in~\cite{kruskals}. Here, we assume that
a graph data structure with Kruskal's algorithm is used in a security
setting where the graph nodes are public and the structure of the graph
(the connection of nodes) is secret.

\noindent\textit{Inputs}.
The input generation for Kruskal's algorithm is based on the domain
knowledge of this problem.
Given that the structure of graphs is secret, the defender
constructs 4,800 graphs as the following.
The defender considers 120 different graph structures
from the interval between
a spanning tree (${n - 1}$) and a complete graph
(${n \times (n-1)/2}$). For each structure,
the number of nodes (n) varies from 2 to 200 and
the number of edges is determined based on
the structure of the graph.
For example, if the structure of a graph
is a spanning tree, the number of edges varies from 1 to 199.

\noindent\textit{Side Channel Discovery}.
For each graph structure,
\toolname fits timing functions that are from the number of nodes
to the execution time.
Then, \toolname applies the clustering algorithm and discovers $20$
clusters with ${\varepsilon_{\scalebox{.6}{0,1}}=0.3}$.
The presence of the $20$ clusters indicate the possibility of
information leaks about the graph structure.

\noindent\textit{Side Channel Explanation}.
The next step is to find out what properties of program internals
are leaking and establish the facts about the leaks.
We obtain program internal features and apply decision tree
algorithms on the set of features for different secret values.
The model shows the number of calls to the \texttt{compareTo}
method distinguish different clusters. This
indicates the sorting algorithm in the MST calculation that
depends on the number of edges is the cause of different observations.
An eavesdropper can
use the side channel to guess whether the graph is a sparse
graph or a dense graph.
The overall analysis time takes about 4 (s).

%
%

\noindent\textbf{J) Collab.}
Collab is a scheduling application that allows users to
create a new event and modify existing ones~\cite{collab}.
Users can apply $add$, $commit$, and $search$
operations on events.
An audit event is a secret, while other events are public.

\noindent\textit{Inputs}.
The defender considers 176 users in the system,
each with either zero or one audit events. The public inputs
are the operations performed on the public events of users.

\noindent\textit{Side Channel Discovery}.
For each user, \toolname applies
1 to 11 operations randomly from the set of possible operations
on their public events and measure the response times.
\toolname models 176 timing functions, one for each user in the system.
The next step is to find out the classes of observations on these
functions. \toolname discovers only one cluster with a small
tolerance value, and the defender concludes that
no information about the audit events of users is leaking
through timing side channels. The clustering algorithm takes about 1(s).

\section{Related Work}
\label{sec:related}
\noindent\textbf{Noninterference.}
Noninterference notion~\cite{goguen1982security}
has been widely used to enforce confidentiality in various
systems~\cite{sabelfeld2003language,terauchi2005secure,almeida2016verifying}.
Previous works~\cite{DBLP:conf/ccs/ChenFD17,DBLP:conf/icse/nilizadeh} extend
the classical notion of noninterference with relaxed notions called
$\varepsilon$-bounded noninterference.
We adopt the well-established noninterference definition to
the functional setting with various noise models.

\noindent\textbf{Static Analysis for side channels.}
Various works~\cite{DBLP:conf/ccs/ChenFD17,antonopoulos2017decomposition,
doychev2015cacheaudit,wang2017cached,TOSEM19} use static analysis
for side-channel detections.
Chen et al.~\cite{DBLP:conf/ccs/ChenFD17} casts the noninterference property
as 2-safety property~\cite{barthe2004secure}
and uses Cartesian Hoare Logic~\cite{barthe2004secure} equipped with taint
analysis~\cite{livshits2005finding} to detect side channels.
These static techniques rely on the taint analysis that is computationally
difficult for real-world Java applications.
The work~\cite{landman2017challenges} reported that 78\% of 461 open-source
Java projects use dynamic features such as
reflections that are problematic for static analysis.
We use dynamic analysis that handles the reflections
and scales well for the real-world applications.

\noindent\textbf{Dynamic Analysis for side channels.}
Dynamic analysis has been used for side-channel detections
~\cite{milushev2012noninterference,DBLP:conf/icse/nilizadeh,profit2019,DBLP:conf/cav/Tizpaz-NiariC019,RV19}.
We compared our technique to DifFuzz~\cite{DBLP:conf/icse/nilizadeh}
in Section~\ref{sec:experiment}.
Profit~\cite{profit2019} considers a black-box model of programs
and study information leaks through network traffics. It first aligns
different traces of packets to identify phases in the application.
Then, it extracts packet-level features such as the time differences
between two packets. Finally,
it uses Shannon entropy to quantify information leaks related to each
feature and provide a ranking of features based on the amounts of leaks.
The trace alignment in Profit is analogous to clustering
in our technique to align traces of different secrets with similar timing
profiles. Similarly, the packet-level features are analogous to extracting
program internal features. The most important difference is the use-case:
our model of systems is white-box
and useful for defenders who have access to the systems.
We consider the variations in both secret and public inputs, while the
variations in Profit~\cite{profit2019} is mostly related to secrets.
While Profit could quantify information leaks, it can't
find out what properties of secrets are leaking. We utilize
program internal features and classifiers to localize code regions
correlated with different observations and establish facts about
leaking properties.

\noindent\textbf{Side-channel Models.}
Chosen-message threats~\cite{kopf2009provably} where
attackers can control public inputs are recently extended for different
attack models~\cite{phan2017synthesis,bang2016string,pasareanu2016multi}.
Phan et al.~\cite{phan2017synthesis} consider synthesizing
adaptive side channels where in each step of the attack,
the attacker chooses the best
public input that maximizes the amount of information leaks.
In our known-message threat model~\cite{kopf2009provably},
however, the attacker only knows public inputs
and may not control them to choose ideal public inputs.
Many related works~\cite{phan2017synthesis,pasareanu2016multi,
DBLP:conf/icse/nilizadeh,DBLP:conf/ccs/ChenFD17} assume that
the observations such as execution times are precise and
apply abstractions such as the number of executed instructions.
However, we support both realistic settings where the observations
are noisy timing measurements and abstractions.

\noindent\textbf{Quantification of information leaks.}
The amount of leaks can be estimated based on quantitative
information flow~\cite{KB07,smith2009foundations,backes2009automatic,KS10,chothia2013tool}.
Smith~\cite{smith2009foundations} defines min-entropy measure
to quantify information leaks. With the assumption that the secret inputs
are uniformly distributed and the program is deterministic,
Smith~\cite{smith2009foundations} shows that the amount of information
leaked based on the min-entropy is $log_2 |L|$ where $L$ is
the classes of observations over the secret set.
Our clustering algorithms can exploit the min-entropy measure defined by Smith~\cite{smith2009foundations} and give lower-bounds on the information leaks.

\noindent\textbf{Localization of vulnerable code fragments.}
Machine learning techniques have been used to detect and
pinpoint culprit codes
~\cite{tizpaz2017discriminating,aaai18,song2014statistical}.
Tizpaz-Niari et al.~\cite{aaai18} consider performance issues in
Java applications. They cluster the execution time of applications and
then explain what program properties distinguish
different functional clusters. The work~\cite{aaai18} is limited to
linear functions (as it needs to discover functions), while
ours supports arbitrary timing functions over public inputs.
In our security context, the program internal features
can be functional. We use an extension of the decision tree
algorithm in~\cite{aaai18} to interpret different clusters.
Symbolic executions have also been used
to find vulnerable fragments~\cite{wang2017cached,guo2018adversarial,wu2018eliminating}.
Richer explanatory models are a unique aspect of our work.
Our decision trees pinpoint basic blocks, contributing
to different observations, as functions of public inputs.

\section{Threat to Validity}
\label{sec:threat}
\noindent\textbf{Overheads in Dynamic Analysis.}
We proposed a dynamic analysis approach to analyze functional
side channels.
Dynamic analysis often scales well to large applications.
However, as compared with static analysis, they present additional overheads
such as time required to discover variegated inputs and time needed for
data collection. 

\noindent\textbf{Functional Regression and Order on Input Data.}
Our approach assumes the existence of an order over the public inputs
to model timing functions.
While such an order is natural for numerical variables, it may require ingenuity
to define a suitable order for data types such as \texttt{strings}
and \texttt{BitStream}.
While our approach can work with any arbitrary user-defined ordering, often a
suitable ordering can significantly improve the simplicity of the timing functions in
the functional regression process.
For instance, compare Figures~\ref{fig:illust-clust} (d)
and Figure~\ref{fig:GF-DT} (d).
Both of these applications model the leaks of set bits with different orders on
the public inputs.
Our approach captures the clusters in both examples, despite the ordering in
Figure~\ref{fig:GF-DT} (d) results in simpler functions.
In practice, we restrict the functions explored in our regression to the class
of basis-splines (B-splines).
These models are parameterized by a given degree to model timing functions, and
regression is more efficient with low-degree splines.
In the case of higher-order target functions, we propose Gaussian
Processes as an alternative to model timing functions.

\noindent\textbf{Use of Decision Trees.}
The proposed decision tree models for discriminant learning partition the space
of auxiliary features into hyper-rectangular sub-spaces.
More expressive models, such as graph models, can be employed to learn
richer classes of discriminants.
However, we posit that simpler models like decision trees provide better
interpretability.
Another simplifying assumption in our approach is to model auxiliary features as
functional attributes and map them to categorical labels.
A more general approach would be to map the functions to numerical values and allow
decision tree algorithms explore the space of features to identify
suitable partitions.
Further analysis of such mapping is left for future work.

\noindent\textbf{Input Generations.}
Our approach requires a diverse set of inputs either given by users or generated
automatically using the fuzzer.
For instance, we used \toolname to generate inputs for the \textit{Regex} case
study, while we use the inputs relevant to known vulnerabilities from DARPA STAC
program for \textit{SanpBuddy}.
The quality of the debugging significantly depends on the presence of
functional side channels in the given input set.
Our fuzzing approach relies on heuristics to generate a diverse
set of inputs, similar to existing evolutionary fuzzers.

\noindent\textbf{Comparison with DifFuzz.}
We compared our approach against DifFuzz~\cite{DBLP:conf/icse/nilizadeh} in
Section~\ref{sec:experiment}.
We chose DifFuzz as an example of dynamic analysis
tool with the point-wise definition of noninterference. We showed that
the functional definition of noninterference gives a realistic sense of security.
Since the clustering as the main tool for finding classes of observations took
place after the input generations, the comparison may not evaluate the
fuzzing engines accurately. We left combining fuzzing and clustering
to detect the number of clusters during the input generations for future work.

\noindent\textbf{Timing Measurements.}
The time observations in our case studies are measured on the
NUC machine (see Section~\ref{sec:environment}) to allow for higher
precision in time and network observations.
To further mitigate the effects of environmental factors such as Garbage
Collections on timing measurements, we take the average of such measurements
over multiple samples. In addition, we turned off JIT compiler for a better
precision.

\section{Conclusion and Future Work}
\label{sec:conc}
We focused on the known-message setting under the assumption that
secret inputs are less volatile than public inputs.
In this setting, the observations appear as timing
functions. We propose a notion of noninterference in the functional setting
and show that it allows defenders
to detect side channels using functional data
clustering. We propose decision tree algorithms to pinpoint locations
in the program that contribute to the side channels.
Our tool \toolname scales well for large real-world
applications and aids debuggers to identify vulnerable fragments
in such applications.

This work opens potential promising directions for future work.
One direction is to combine the fuzzer with clustering that
can directly estimate the number of distinguishable observations
during the input generations. In this case, the objective is to find $n$ secret values
and $m$ public values and maximize the number of distinguishable
clusters in timing observations.
Another direction is to study the potential timing side
channels for machine learning applications. Given a learning problem with
$n$ samples and $m$ features as public inputs,
the feasibility of leaking (hyper-)parameter~\cite{wang2018stealing}
of machine learning models via timing side channels
is a relevant and challenging problem.

\noindent\textit{Acknowledgements.}
The authors would like to thank the anonymous reviewers for their valuable comments
to improve our paper. This research was supported by DARPA under
agreement FA8750-15-2-0096.


\bibliographystyle{IEEEtran}
\bibliography{papers}


\end{document}